%% file: main.tex
\documentclass[a4paper,fleqn]{cas-dc}

\PassOptionsToPackage{final}{changes}

\usepackage{graphicx}
\usepackage{subcaption}

\usepackage[sort&compress,numbers]{natbib}
\usepackage{chngcntr} 
\input{preamble.tex} 

\begin{document}
\let\WriteBookmarks\relax
\def\floatpagepagefraction{1}
\def\textpagefraction{.001}

\shorttitle{Let it shine: Autofluorescence of Pap-stain for AI-based Cancer Detection}    
\shortauthors{Lian, Lindblad,\ldots Sladoje}  

\title[mode = title]{ Let it shine: Autofluorescence of Papanicolaou-stain improves AI-based cytological oral cancer detection  }  


\definechangesauthor[color=Violet]{WL}
\definechangesauthor[color=BrickRed]{NS}
\definechangesauthor[color=NavyBlue]{JL}

\author[1]{Wenyi Lian}[orcid=0009-0000-3069-5366]
\ead{wenyi.lian.7322@student.uu.se}


\author[1]{Joakim Lindblad}[orcid=0000-0001-7312-8222]
\ead{joakim.lindblad@it.uu.se}

\author[2,3]{Christina {Runow Stark}}
\author[2]{Jan-Micha\'el Hirsch}

\author[1]{Nata\v{s}a  Sladoje}[orcid=0000-0002-6041-6310]
\cormark[1] 
\ead{natasa.sladoje@it.uu.se}

\affiliation[1]{organization={Centre for Image Analysis, Department of Information Technology, Uppsala University},
            city={Uppsala},
            country={Sweden}}

\affiliation[2]{organization={Department of Surgical Sciences, Uppsala University},
            city={Uppsala},
            country={Sweden}}

\affiliation[3]{organization={Folktandvården, Region Uppsala},
            city={Uppsala},
            country={Sweden}}

\cortext[1]{Corresponding author}

\begin{abstract}
    \noindent \textit{Background and objectives:} Oral cancer is a global health challenge. The disease can be successfully treated if detected early, but the survival rate drops significantly for late stage cases. There is a growing interest in a shift from the current standard of invasive and time-consuming tissue sampling and histological examination, toward non-invasive brush biopsies and cytological examination, facilitating continued risk group monitoring. For cost effective and accurate cytological analysis, there is a great need for reliable computer-assisted data-driven approaches. However, \replaced{infeasibility}{lack of} of accurate cell-level annotation\deleted{s} hinders model performance, and limits evaluation and interpretation of the results. This study aims to improve AI-based oral cancer detection by introducing additional information through multimodal imaging and deep multimodal information fusion.
    
    \noindent \textit{Methods:}  We combine brightfield and fluorescence whole slide microscopy imaging to analyze Papanicolaou-stained liquid-based cytology slides of brush biopsies collected from both healthy and cancer patients. Given the challenge of detailed cytological annotations, we utilize a weakly supervised deep learning approach only relying on patient-level labels.  We evaluate various multimodal information fusion strategies, including early, late, and three recent intermediate fusion methods.

    \noindent \textit{Results:} Our experiments demonstrate that: (i) there is substantial diagnostic information to gain from fluorescence imaging of Papanicolaou-stained cytological samples, (ii) multimodal information fusion improves classification performance and cancer detection accuracy, compared to single-modality approaches. Intermediate fusion emerges as the leading method among the studied approaches. Specifically, the Co-Attention Fusion Network (CAFNet) model achieves impressive results, with an F1 score of 83.34\% and an accuracy of 91.79\% \added{at cell level}, surpassing human performance on the task.  Additional tests highlight the importance of accurate image registration to maximize the benefits of the multimodal analysis.

    \noindent \textit{Conclusion:} This study advances the field of cytopathology by integrating deep learning methods, multimodal imaging and information fusion to enhance non-invasive early detection of oral cancer. Our approach not only improves diagnostic accuracy, but also allows an efficient, yet uncomplicated, clinical workflow. The developed pipeline has potential applications in other cytological analysis settings. We provide a validated open-source analysis framework and share a unique multimodal oral cancer dataset to support further research and innovation.
\end{abstract}





\maketitle

\section{Introduction}
\label{sec:intro}

Cancers of the oral cavity and oropharynx are among the most common malignancies in the world. The \textit{GLOBOCAN} database reports that the annual incidence of lip, oral and oropharyngeal cancer in 2022 reached 500,000, and a mortality of 240,000~\cite{globocan2022}.  
Early detection enables timely treatment, thereby substantially improving survival. 
The average 5-year survival rate in the United States is 69\% (2014--2020), but this is highly dependent on when the cancer is found. Only 26\% of malignancies are diagnosed while at a local stage, when the 5-year relative survival is as high as 87.5\%. Instead, the cancer is often detected late, when it has metastasized, in which case the 5-year survival rate drops to 37.8\%~\cite{seerProgram}.

The current medical standard for setting a diagnosis of Oral Cancer (OC) is based on histological examination of a tissue biopsy sample 
obtained from the suspected tumor site; a time consuming, resource-demanding, and painful procedure. With faster, cheaper, more accessible and non-invasive approaches,  early detection, treatment and prognosis of oral and oropharyngeal cancer could be significantly improved~\cite{hirsch2023paradigm,haj2024early}.

An appealing alternative to histological analysis is  {\it cytological analysis of brush samples} (exfoliative cytology) from the oral cavity of the patients. Brush biopsies can painlessly, and relatively easily~\cite{EDMAN2023} be taken during regular visits to a dentist, when suspicious changes in the oral mucosa are observed. The brush samples are then undergoing cytological analysis, commonly performed by a human expert who, using brightfield microscopy (BF), examines Papanicolaou (Pap)-stained cells on a prepared slide. A trained cytologist can detect abnormalities in samples acquired from patients with malignancy, however such visual assessment is both very difficult and very time-consuming. As reported in~\cite{sukegawa2020clinical}, sensitivity and specificity reached by human experts {\it on the patient-level diagnosis} are $80\%$ and $86\%$, respectively, in liquid-based OC cytology screening (assuming oral high-grade squamous intraepithelial lesion and squamous cell carcinoma as positive). Considering the difficulty to pinpoint very subtle malignant changes in a few, among perhaps 100,000, cells on a glass, there is a great desire for ways to improve the diagnostic accuracy.

Even though an increasing interest in utilizing AI-based decision support for cancer detection in cytology is evident~\cite{LANDAU2019230, KIM202497, kavyashree2024systematic}, AI-based methods for OC detection are still scarce.  Existing results are promising and have confirmed that Deep Neural Networks (DNN) can be trained on Whole Slide Images (WSIs) of Pap-stained cytological samples, to detect OC~\cite{wieslander2017deep, lu2020deep, koriakina2024PLOS}.

A well recognized challenge for the development of AI-based methods in cytology is the severe lack of detailed and reliable annotations. It is not feasible for a cytologist to label each individual cell of a WSI as malignant or healthy, both because of the very large number of cells in each sample (5,000--150,000 cells), and because of the extreme difficulty of the task, where often only very minute differences separate the classes\deleted{)}. Therefore, supervised training of DNNs to detect malignancy on the cell level is not realizable. Instead, per-patient labeling is typically used: a label is assigned to the patient's cytological sample, based on histological analysis of a tissue sample acquired at the same location, or (\replaced{if available}{more reliably}) future patient outcome. This type of labeling leads to a weakly supervised learning problem, imposing challenges in training/learning as well as for reliable performance evaluation of developed automated systems.    

We aim to increase the amount of information available for automated detection of OC without requiring extensive manual effort invested in per-cell annotations. The latest trends in cancer studies, and life sciences in general, extensively explore the potential of heterogeneous and complementary information about specimens, acquired through diverse imaging (and other) systems and combined in a subsequent correlative analysis of such \textit{multimodal} data. These trends are supported by recent AI-based analysis approaches, which offer efficient multimodal information extraction and fusion~\cite{LIPKOVA2022, BOEHM2022,Vollmer2024,ZITNIK201971}. In this study, we evaluate if multimodal cytological image information can be used to improve OC detection.

Multimodal approaches have been explored for OC detection, confirming advantages of combining complementary information from multiple (macroscopic and microscopic) scales~\cite{PAL2020}, or different stainings~\cite{remmerbach2009toward}.  
However, such approaches typically require advanced sample preparations, which adds to the complexity of imaging. In addition, the correlated analysis most often requires (some type of) multimodal image alignment to establish correspondence between the acquired multimodal data. 

To ensure high clinical relevance and applicability of our study, we investigate if an advancement in OC detection can be reached through the correlative analysis of Pap-stained cells from brush biopsies imaged by both BF and fluorescence microscopy (FL) on exactly the same sample, avoiding any additional staining, and simplifying the required multimodal image alignment. 
Inspiring studies on fluorescence emitted by Pap-stained samples for improved urothelial cancer detection~\cite{steenkeste2007ex}, and for improved detection of inflammation and bacterial infections of respiratory system~\cite{WEHLE1991,KUPPER1995,HETTLICH1998} exist, however no similar study has previously been performed in OC detection.

The main contributions of our work are:
(i) We propose an approach based on multimodal imaging of liquid-based cytology (LBC) prepared slides from brush biopsies, combined with AI-based cytological analysis and multimodal information fusion, for improved OC detection;
(ii) We propose to combine fluorescence and brightfield WSI, and confirm that this simple and fast multimodal imaging, which requires only one staining, enables OC detection that outperforms approaches based on either of the individual modalities; 
(iii) We evaluate different deep learning strategies for multimodal information fusion: early, late and intermediate, observing that the intermediate fusion performs best on this task; 
(iv) We device an efficient multimodal registration pipeline for highly accurate registration of the FL and BF images and confirm, through experimental evaluation, that this is crucial for high OC detection performance; 
(v) We collect and share a multimodal OC dataset with $766{,}565$ aligned BF and FL image pairs of extracted cells from cytological WSIs: \url{In_Prepraration}\footnote{Will be completed in time for article publication.};
(vi) We contribute to the development of an accurate and reliable, and therefore clinically relevant, AI-based support system for early OC detection. 

To facilitate reproducibility, we share our complete implementation and evaluation framework as open source: \url{https://github.com/MIDA-group/OralCancerMultimodal}. 

\section{Related Work}
\label{sec:rw}

Our long-term focus is on the development of an efficient and reliable automated system for early OC detection based on cytological analysis of brush biopsies.  In this work, we propose to utilize multimodal image information fusion to increase the informative content extracted from patients' cytological samples by an AI-based cancer detection system. In this section we present previous studies and the existing results that we build on. They cover most relevant aspects of our proposed approach: (i) state-of-the-art in AI-based OC detection by cytological analysis; (ii) experiences of use of multimodal image information in OC detection; (iii) relevance of fluorescence of Pap-stained cells in cancer studies; (iv) state-of-the-art AI-based systems for (biological/biomedical) image information fusion.

\subsection*{AI-supported cytology for OC detection}

A survey on AI-methods in cytology~\cite{LANDAU2019230} from 2019 does not mention any studies on AI-based OC detection among the increasing number of cancer types detected by emerging AI-based approaches. A 2022 survey on AI-approaches for OC detection~\cite{ALRAWI2022436} 
does not report any methods applicable to microscopy images of brush biopsies; 
primary focus is on histological analysis or analysis of photographic images of the oral cavity. A 2021 survey on AI-supported early detection of OC~\cite{GARCIA2021} does report on several approaches developed to support exfoliative cytology (including analysis of brush biopsies). Methods mainly rely on shallow machine learning (i.e., manually extracted features, often in combination with Support Vector Machines), and confirm feasibility of AI-based early OC detection from cytological samples, calling for further research and method development. The work presented in~\cite{wieslander2017deep} evaluates DNNs for the task, and is, as such, a forerunner to our here presented work. \added{This early study does not consider WSIs, but only manually selected regions of cells, and does not include any automated cell detection.}

Whole slide imaging is increasingly being used in cytology. 
A WSI of a cytological slide typically contains tens of thousands of cells. For a sample from a healthy patient, all collected cells are healthy, whereas for a sample from a patient with cancer, some of the cells are malignant (while the majority typically are healthy also in this case). Reliable individual cell-annotation is not feasible, due to the large number of cells and the difficulty of the task. Cytologists, when making a {\it patient-level} assessment (i.e., assessment of the whole slide, by aggregating information of identified characteristic properties of malignant cells, or absence of such), reach a sensitivity and specificity of the patient diagnosis of around $80\%$ and $86\%$, respectively~\cite{sukegawa2020clinical}.

However, the unit of analysis in cytology is the cell and a patient diagnosis is determined based on the presence or absence of malignant cells in a sample.  A natural approach in WSI analysis in cytology is, therefore, to extract image patches of a fixed size, each containing a centered in-focus cell nucleus, and to process these by a deep learning classification system. An illustrative example of such extracted cell-patches from a WSI (in two modalities) is shown in Fig.~\ref{fig:patch_extract}.  To reach reliable training and validation of such classification models, researchers rely on patient-level annotations, where labels are derived from histological analysis of biopsies, or (if available) from information about future patient outcome. True labels of the individual cells are not known, but are considered to coincide with the label of the sample they belong to. (In other words, the cell label does not indicated malignancy of that cell, but whether the cell originates from a patient with cancer or not.) Consequently, strategies suitable for learning from weak labels, and in particular, Multiple Instance Learning (MIL) approaches~\cite{CARBONNEAU2018329,cheplygina2019not}, are considered. At the same time, the number of malignant cells in a sample is typically much smaller than the total number of cells, which constitutes yet another challenge -- MIL on imbalanced training data with very few key instances (malignant cells that determine the diagnosis).
Furthermore, while the number of cells in a sample is typically very large, the number of samples (patients) available in studies is often rather low, which is a serious limitation for AI-model development.

Existing studies confirm that an AI-system can be trained on such weakly labeled \added{monomodal} cytology data to detect subtle changes in cell morphology caused by malignancy~\cite{lu2020deep,koriakina2021effect}. The CNN-based model proposed in~\cite{lu2020deep} predicts\added{, based on BF data,} the patient-level label by aggregating the corresponding cell-instance predictions; it is evaluated in terms of F1-score (harmonic mean of precision and recall), a suitable performance measure for imbalanced data. An alternative, based on Attention-Based MIL (ABMIL)~\cite{ilse2018attention}, and modified for efficient use with very large bags is proposed in~\cite{koriakina2021effect,andersson2022end}. A comparative study~\cite{koriakina2024PLOS}, indicates that the approach proposed in~\cite{lu2020deep} (referred to as Single Instance Learning (SIL) approach) delivers better performance in OC detection than ABMIL with sampling~\cite{koriakina2021effect, koriakina2024PLOS}, while being less complex and less memory-demanding. It also exhibits better ability to detect key instances, in particular for situations where the ratio of key instances in positive samples is very low\added{ (a case which is desired to handle for early cancer detection)}. Our here presented study therefore builds on the SIL approach as in~\cite{lu2020deep}, adjusting it to the multimodal setting.

\subsection*{Multimodal imaging and image analysis for OC detection }

Multimodal methods for OC detection are most often developed for tissue analysis (of histological biopsies or in vivo); examples include a combination of macroscopic biochemical imaging of fluorescence lifetime imaging (FLIM) and subcellular morphological imaging of reflectance confocal microscopy (RCM)~\cite{MALIK2016}, in vivo assessment by macroscopic white-light imaging of autofluorescence followed by high-resolution microendoscopy of identified regions ~\cite{YANG2019}, and widefield fluorescence (WF) imaging with non-linear optical microscopy (Multi-Photon and Second Harmonic Generation)~\cite{PAL2020}.

Studies that confirm an increase of informative content in oral brush biopsies from sequential staining of the samples include an early example presented in~\cite{remmerbach2009toward}, where  Pap, Feulgen, and Silver nitrate staining were combined in (manually performed) correlative analysis. This study, which does not rely on any specialized integrated multimodal imaging system, is relevant for our use-case since it confirms an information gain from multimodal imaging; however we aim to avoid re-staining of the samples as well as to automate the process of extraction and fusion of information.

\subsection*{Fluorescence microscopy of Pap-stained cytological slides}

The studies\deleted{ presented in}~\cite{steenkeste2007ex,ALSIBAI2020}  
show that FL imaging of
Pap-stained urothelial cells \replaced{reveals information which is relevant for}{allows} the detection of early neoplastic lesions. The distributions of fluorescence for tumoral urothelial cells and normal cells appear rather different (the former being characterized by a
perimembrane fluorescence localization, and the latter exhibiting an intracellular fluorescence). The\replaced{se}{ir} results, relying on the analysis of photophysical properties of the different dyes used for the Pap staining, encourage utilization of FL imaging, \replaced{while}{without} changing neither the sampling of the specimens nor the Pap-staining protocol. 
Further analysis~\cite{GUTIERREZ2022}, provided some insight in the underlying biological phenomena behind the peri-membrane fluorescence. 

Several other studies have explored the advantages of FL of Pap-stained samples in contexts different from cancer detection. For detection of mycobacterial species in lymph node fine-needle aspiration specimens~\cite{WRIGHT2004}, it is observed that the fluorescence information, complementary to the standard brightfield microscopy, improves specificity of the analysis, with a minimal additional investment. Other related results include cytological diagnosis of Pneumocystis carinii in bronchoalveolar lavage specimens~\cite{WEHLE1991}, detection of Aspergilius infections~\cite{HETTLICH1998}, detection of oral Candida~\cite{titinchi2011diagnostic} and the diagnosis of Mycobacterium kansasi tuberculosis~\cite{KUPPER1995}, all indicating informative content of the (auto-)fluorescence of Pap-stained samples. 

\replaced{Inspired by}{We build on} these results, 
\replaced{we}{and} acquire both BF and FL images of Pap-stained LBC oral cytological slides\replaced{.}{,} 
\added{We aim}
to enable AI-based multimodal information extraction and integration, towards improved early OC detection.

\begin{figure*}
    \centering
    \includegraphics[width=0.78\linewidth]{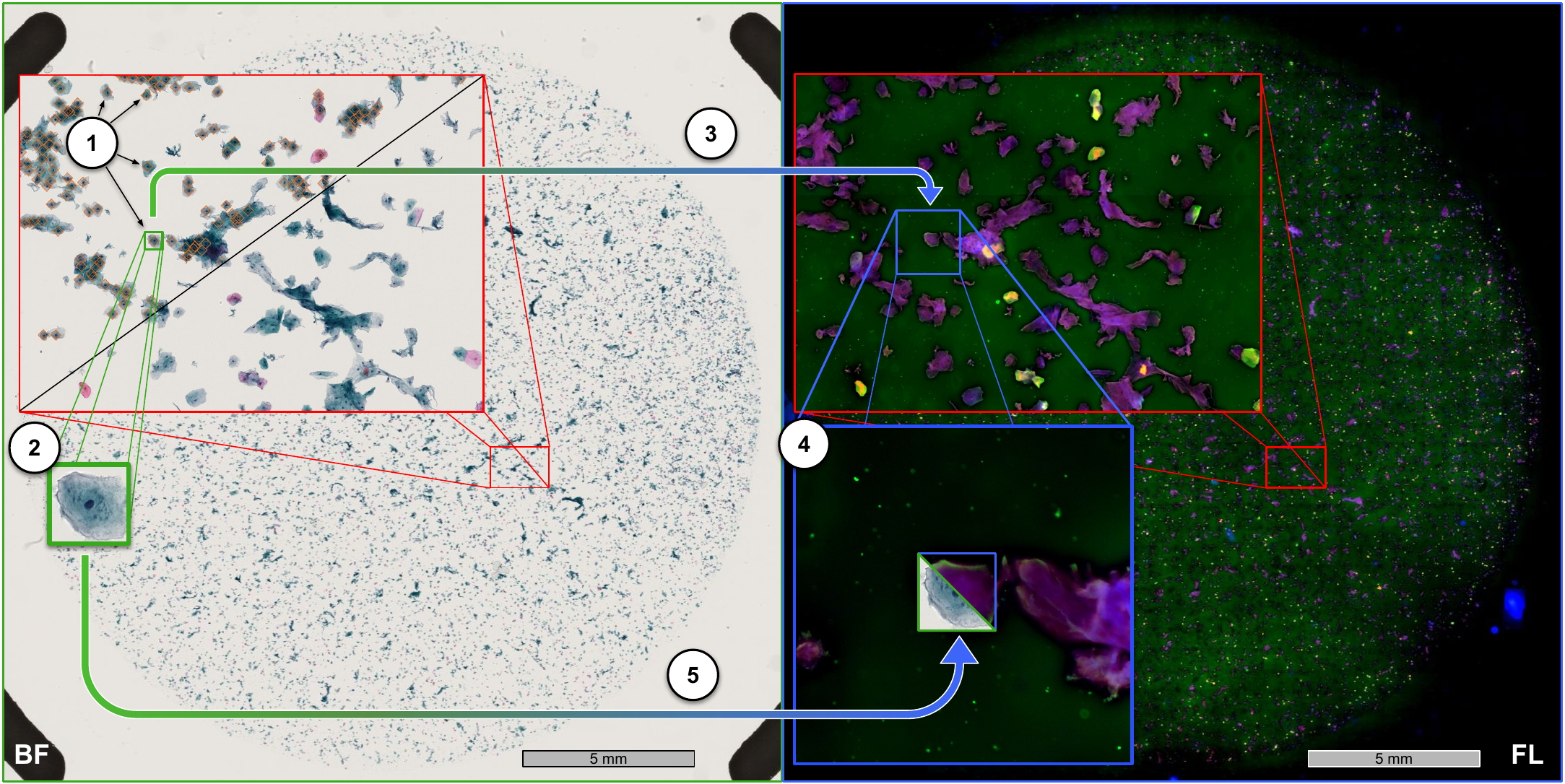}
    \includegraphics[width=0.21\linewidth]{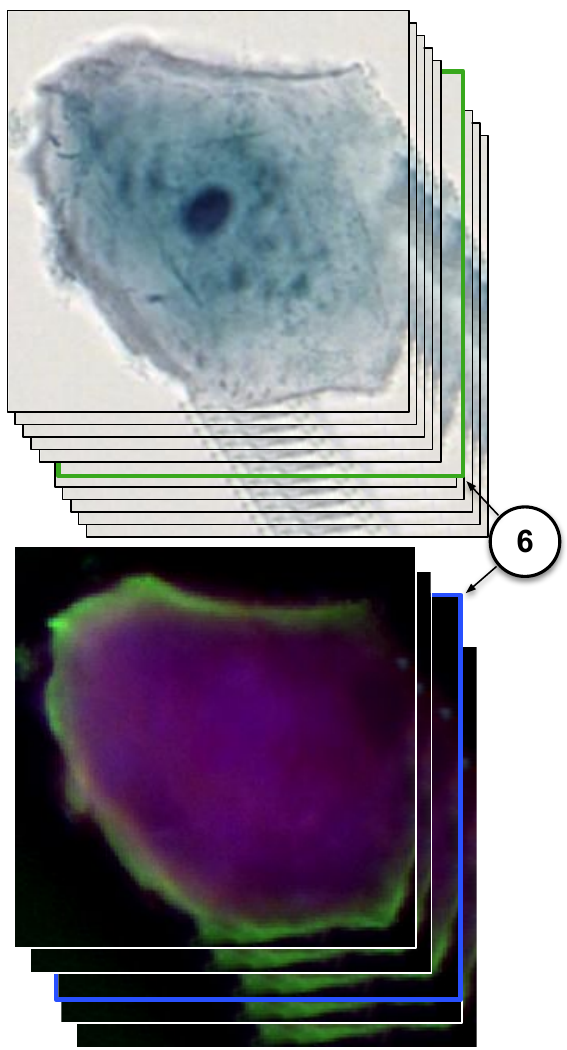}
    \caption{Illustration of the aligned patch extraction procedure, starting from a rigidly aligned pair of BF and FL WSIs. \textbf{1}. Nucleus detection in BF (orange lozenges; 31,336 detected in this slide); \textbf{2}. Patch extraction $256\!\times\!256\,$px in BF (green square); \textbf{3}. Rigid mapping of nucleus position into FL image; \textbf{4}. Large $768\!\times\!768\,$px patch extraction (larger blue square); \textbf{5}. Translation only CMIF-based registration for pixel perfect $256\!\times\!256\,$px patch extraction in FL image; \textbf{6}. Selection of best the focus level in BF (11 levels) and FL (5 levels). Only three out of the four FL channels (emission wavelengths 668,517,465) are visualized (as RGB).}
    \label{fig:patch_extract}
\end{figure*}

\subsection*{AI-based multimodal information fusion} 

Integration of complementary information from diverse data sources allows more reliable cancer diagnosis. However, to perform manual/visual correlative analysis across diverse data modalities becomes a very complex task, calling for support by AI-based information fusion and analysis methods~\cite{LIPKOVA2022,BOEHM2022}.
Different strategies are considered in a variety of studies~\cite{ramachandram2017deep, ZITNIK201971}.  {\it Early fusion} utilizes one model to process the combined information from all modalities. The combined input can be created by,  e.g., concatenation of the input vectors, or their element-wise sums or products. {\it Late fusion}, on the other hand, integrates information from different modalities at the final decision level, after a separate model is trained for each modality. The aggregation of the  models' predictions can be performed by e.g., averaging, majority voting, or through a learned model. 



To allow a 
gradual influence of the multimodal information on the extraction of features from individual modalities, \replaced{a number of}{different} strategies for {\it intermediate fusion} have been suggested. The different methods exploit the multimodal information in different ways, allowing the intermediate integrated content (features) to backpropagate and influence, to a higher or lower extent, the overall process and its output. Relevant examples include~\cite{joze2020mmtm,bi2020multi,he2023co}. 
%

In our here presented study we evaluate \replaced{early, late, and intermediate multimodal information fusion}{three recent} strategies\deleted{ for multimodal information fusion}, to reach improved OC detection from BF and FL Pap-stained microscopy images.

\section{Multimodal Oral Cancer Dataset}
\label{sec:data}

For the purpose of this study we create a multimodal OC dataset consisting of accurately aligned  BF and FL images (n=766,565) of Pap-stained in-focus cells acquired from healthy individuals and patients with histologically confirmed OC. The procedure employed to create the dataset is illustrated in Fig.~\ref{fig:patch_extract} and described in detail below.

\subsubsection*{Sample acquisition}

We collected
LBC prepared Pap-stained slides of brush-sampled cells from the oral cavity of 19 adult patients. Positive samples (n=8) were obtained from the anatomical location of histopathologically verified oral squamous cell carcinoma lesions. 
Out of the negative samples (n=11), two were diagnosed as healthy based on histopathological analysis\deleted{ (similar as for the positive samples)}, whereas the majority of negative samples (n=9) were acquired from healthy adults 
without an accompanying histological analysis due to the invasive nature of the associated tissue biopsy.

Samples were collected (Folktandvården, Dept. of Orofacial Medicine, Södersjukhuset, Region Stockholm, Sweden) with a Cytobrush Plus GT (Medscand Medical, Cooper Surgical company, USA) for cytological assessment. The brush was rubbed against the oral mucosa for approximately 30\,s and then immediately placed in a vial of PreservCyt\texttrademark{} transport medium (Hologic, Inc. USA), spun around the walls of the vial for approximately 10\,s and then removed. Collected specimens were processed according to the standard procedure at the Department of Pathology and Cytology, previously described in~\cite{EDMAN2023}. LBC slides were prepared using a ThinPrep TP5000 processor (Hologic, Inc., Bedford, MA, USA) and stained using a Gemini AS slide Stainer (Thermo Scientific, Gothenburg, Sweden) according to regressive Papanicolaou (Pap) staining technique~\cite{soost1979comparison}.

\subsubsection*{BF and FL microscopy imaging}

We first image the Pap-stained slides under white light using a NanoZoomer S60 digital slide scanner (Clinical Pathology and Cytology, Uppsala University Hospital), $40\times$, 0.75 NA objective, at 11 $z$-offsets (stepping $0.4$\,\textmu m) providing RGB BF WSIs of approximately $100{,}000\times100{,}000$ pixels, $0.23$\,\textmu m/pixel. 

We subsequently acquire FL images of the same Pap-stained slides using a Zeiss Axio Scan Z1 slide scanner (BioVis platform of Uppsala University), $20\times$, 0.8 NA, at 5 $z$-offsets (stepping $1.0$\,\textmu m), excitation wavelengths: $\{353,493,553,653\}$, emission filter wavelengths: $\{465,517,568,668\}$, providing 4-channel FL WSIs of approximately $77{,}000\times77{,}000$ pixels, $0.33$\,\textmu m/pixel.

\subsubsection*{Correction for spatial intensity nonuniformity}
For the FL images, we correct for spatial intensity nonuniformity by, for each channel separately, subtracting a low-pass Gaussian filtered version of the image ($\sigma = 10\%$ of image side length) followed by a linear rescaling of the intensities to the range $[0,1]$, with a saturation of bright pixels which surpass 4 times the 99th percentile.

\subsubsection*{Nucleus detection and focus selection in BF}
We employ a modified version of the method described in Lu \etal~\cite{lu2020deep} to detect cell nuclei in each of the BF WSIs. 
First, images are downsampled by a factor $4\times4$. 
A small (8 convolution layers) regression U-Net~\cite{ronneberger2015u} is trained based on manually marked nucleus locations (each modeled with a small 2D Gaussian kernel, $\sigma=3\,$px) in three slides. (These three slides are not later used in the study.) 
For each slide, nuclei locations are detected at local maxima of height ${>}\,0.5$ in the prediction output. To not miss nuclei in the top or bottom focus-layers, the nucleus detection is run on images from $z$-offsets\deleted{,} $\{-2,0,2\}$\,\textmu m; a detection is registered if found in any of the three $z$-levels. Example output from the nucleus detection step can be seen in Step 1 of Fig.~\ref{fig:patch_extract}.

For each detected nucleus location, $256\!\times\!256$ pixel regions are cut out from each $z$-level at the full image resolution (Step 2 of Fig.~\ref{fig:patch_extract}), and the $z$-level which provides the best focus is selected (BF part of Step 6 of Fig.~\ref{fig:patch_extract}). 
We perform focus selection by using a center weighted (knowing that the nucleus is in the center of the cut out patch) modified Laplacian (LAP2)~\cite{pertuz2013analysis,nayar1994shape}.
\replaced{The method is evaluated on}{On} a separate test set \added{consisting of BF images of 100 nuclei, each acquired at 11 focus levels.} \added{The ``ground truth'' is}
created by asking 8 experts to select the best \replaced{focus level}{of 11 focus-levels} for \replaced{each}{100} nucle\replaced{us}{i} and taking the median of the 8 assigned labels as true best focus\replaced{.}{,} \replaced{The proposed method}{this approach} reaches an accuracy of 93\% \added{on this data set}; \replaced{outperforming}{it outperforms} both the method presented in~\cite{lu2020deep} and the average human expert, the two reaching 84\% and 85.5\% accuracy, respectively, on the same data.


\subsubsection*{Aligned patch extraction from FL data}

To register the FL and BF images, we use the {\it globalign}, global multimodal image registration method proposed in~\cite{ofverstedt2022fast}\replaced{. This method is}{,} based on fast computation of cross-mutual information in the frequency domain. \added{It is shown to reach state-of-the-art performance on a range of modality combinations, including biomedical data.} 

For each pair of FL and BF WSIs, we first estimate a global translation and rotation on a $32\times$ downsampled version, incorporating the known spatial scale factor of $1.472$ between the different modalities. Due to small drift during the scanning and accumulated stitching errors, this slide-level alignment \replaced{does not reach pixel-level accuracy}{is not accurate enough to fully exploit the multimodal information}. We therefore perform a translation-only re-alignment for each individual nucleus. Knowing the global rigid transformation, we compute, for each detected nucleus in the BF image, its corresponding location in the FL image and there extract, at the middle $z$-level, a $768\!\times\!768$ pixel region at the full image resolution (Steps 3 and 4 of Fig.~\ref{fig:patch_extract}). For each extracted region we (again) run {\it globalign} (with known scale and rotation) to find the best match with the corresponding middle $z$-level patch of the BF image at pixel-level accuracy (Step 5 of Fig.~\ref{fig:patch_extract})\replaced{. We then}{, and} extract the corresponding $256\,{\times}\,256$ pixel region from the considered FL patch at the level which provides the best focus, similar as for the BF data (FL part of Step 6 of Fig.~\ref{fig:patch_extract}). 

The contrast in the FL images is sometimes very low, which makes the registration unreliable. To ensure high quality data, we remove the 5\% cells with the lowest FL contrast (based on the used focus measure), as well as cells for which the fine-level registration does not agree with the alignment of neighboring cells (such ``failed registration'' most often coincide with low image contrast).

\subsubsection*{Labeling}
The cells are labeled according to the patient-level labels%
: all cells from the samples originating from patients with a detected malignancy are labeled as ``positive'', whereas all cells from the samples associated with healthy patients are labeled as ``negative''. Note that this labeling does not assign cell-level labels to indicate malignancy, since a majority of cells from patients with malignancy are actually healthy, and only a subset is indeed malignant.

\subsubsection*{Complete dataset}
The above described procedure provides us with in total $766{,}565$ cut-outs, each with an in-focus  cell in the center, well aligned in the two modalities; $599{,}167$ patches are from 11 healthy and $167{,}398$ are from 8 OC patients. We make this collected and accurately aligned multimodal OC dataset available for research purposes: \url{In_Prepraration}\footnote{Will be completed in time for article publication.}.

\section{OC Classification Methods}
\label{sec:method}
\begin{figure*}[t]
    \centering
    \begin{subfigure}[b]{0.1415\textwidth}
      \centering
      \includegraphics[width=0.8\linewidth]{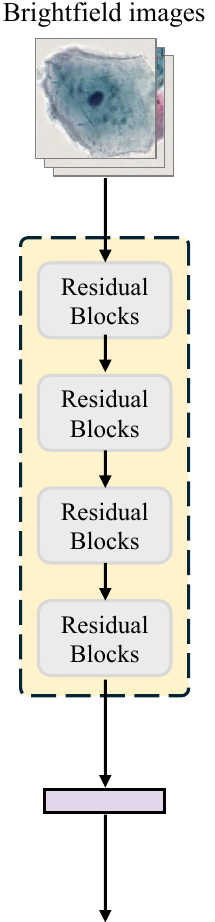}
       \caption{BF only.}
       \label{fig:bf_fusion}
    \end{subfigure}
   \begin{subfigure}[b]{0.1559\textwidth}
      \centering
       \includegraphics[width=0.8\linewidth]{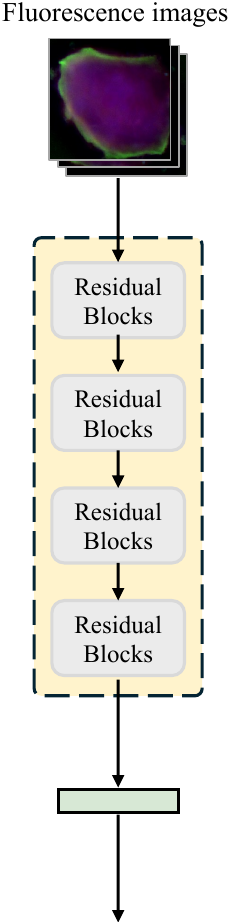}
      \caption{FL only.}
      \label{fig:fl_fusion}
    \end{subfigure}
    \begin{subfigure}[b]{0.315\textwidth}
      \centering
      \includegraphics[width=0.8\linewidth]{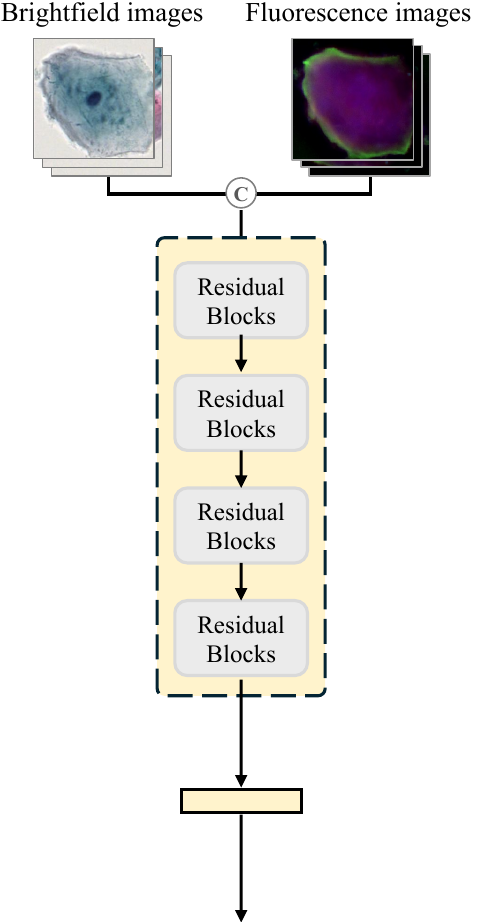}
       \caption{Early fusion.}
       \label{fig:early_fusion}
    \end{subfigure}
   \begin{subfigure}[b]{0.318\textwidth}
      \centering
       \includegraphics[width=0.8\linewidth]{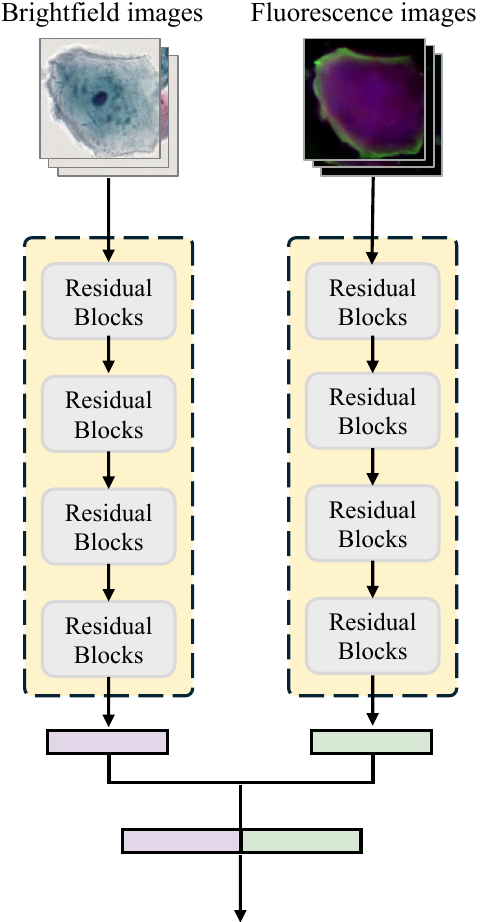}
      \caption{Late fusion.}
      \label{fig:late_fusion}
    \end{subfigure}
    \caption{Classification of microscopy images based on (a) and (b) single-modality; (c) early fusion; (d) late fusion frameworks.}
    \label{fig:fusion_overview}
\end{figure*}

The power of multimodal approaches lies in complementarity of the combined modalities, such that their integrated informative content exceeds the information available from each individual modality alone.
We explore several multimodal fusion strategies and architectures, adjusted to extract and integrate information from BF and FL images of Pap-stained cells, to provide an information-rich content for AI-based OC detection. In the following subsections 
we introduce the main general fusion strategies,  
overview three recent advanced intermediate multimodal fusion networks -- Multimodal Transfer Module (MMTM)~\cite{joze2020mmtm}, Hyper-connected Convolutional Neural Network (HcCNN)~\cite{bi2020multi}, and\deleted{ the} Co-Attention Fusion Network (CAFNet)~\cite{he2023co},    
and finally 
summarize the modifications made to adjust these methods for OC detection from our dataset.

\subsection{Multimodal fusion frameworks}
\label{subsec:frames}

A main challenge in deep learning-based multimodal fusion is how to effectively fuse the different modalities to best use the advantages of the complementary information they bring in, enhancing feature extraction or final representation learning, while still not disrupting the learning process~\cite{baltruvsaitis2018multimodal}. 
Depending on the location of the fusion process in the workflow, the information integration methods are generally categorized into early, late, and intermediate fusion~\cite{ramachandram2017deep}.

\paragraph*{Early fusion} directly merges data sources at the initial processing stage. As depicted in Fig.~\ref{fig:early_fusion},  images from different modalities are merged before being input into the network. This approach allows the exploitation of cross-correlations between different modalities at the pixel level. However, this requires the data to be accurately aligned to reach a unified representation suitable for further joint processing. 

\paragraph*{Late fusion,} as illustrated in Fig.~\ref{fig:late_fusion}, typically involves concatenating high-level features, independently extracted by different (suitably selected) networks from each modality, to perform the final prediction of class labels. 
Late fusion is expected to be more robust than early fusion, since the integration of the high-level features alleviates the need of accurate image alignment. However, due to fusing features only in the final layer, this type of architecture has a much weaker ability to establish, and utilize, correspondence between the different modalities. 

\paragraph*{Intermediate fusion} utilizes features from different levels of the decision model to enhance the model’s ability to capture complex data relationships and to improve accuracy compared to late or early fusion.
Whereas early fusion methods typically processes data in one branch, and late fusion methods comprise two branches (one for each modality), intermediate fusion frameworks often feature an additional (intermediate) ``fusion branch'', to process multi-level fused information in the network. \added{Several approaches have been proposed regarding how to utilize the fusion branch within the network architecture, with the aim to find a most beneficial balance between monomodal feature enhancement and cross-modal information fusion. We evaluate three state-of-the-art deep fusion models, being representative examples of different strategies for intermediate fusion, to find a best performing one for the task of multimodal OC detection.}
\deleted{Some examples follow in the coming subsections.}


\begin{figure*}[t]
  \centering
   \includegraphics[width=1\linewidth]{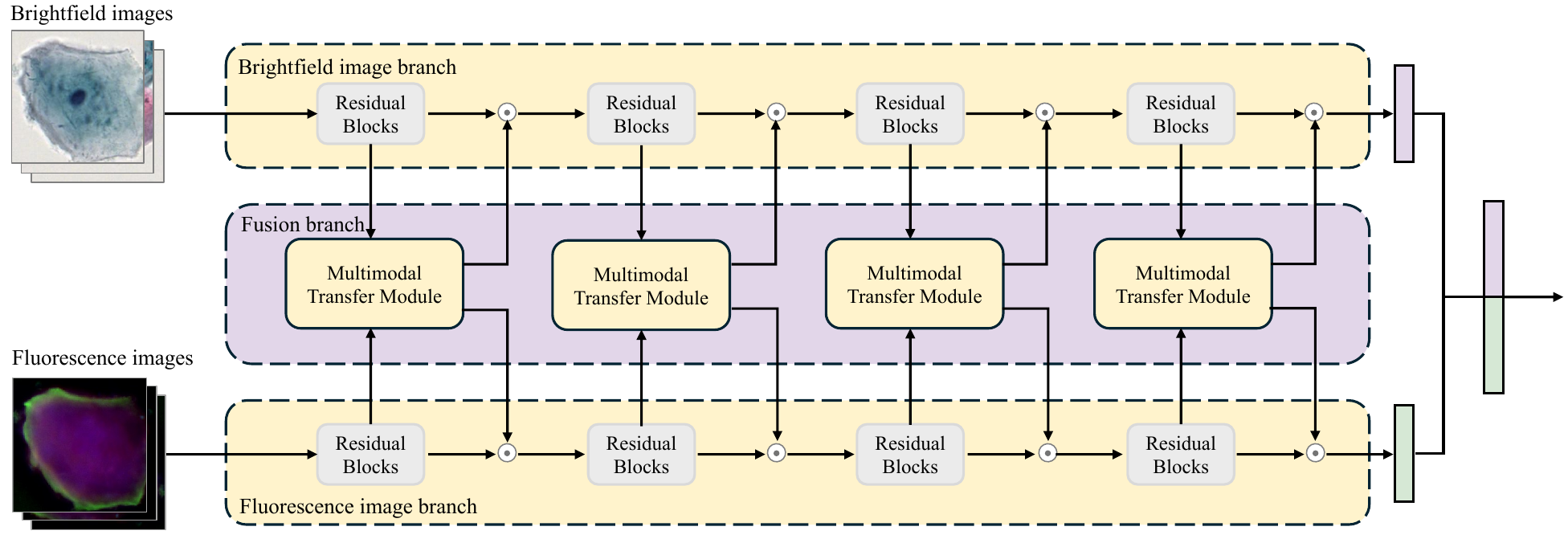}
   \caption{Overview of the Multimodal Transfer Module (MMTM) framework\replaced{,}{;} adapted from~\cite{joze2020mmtm}.}
   \label{fig:mmtm_overview}
\end{figure*}

\subsubsection{MMTM}
\label{subsec:mmtm}

\begin{figure}[t]
  \centering
   \includegraphics[width=1\linewidth]{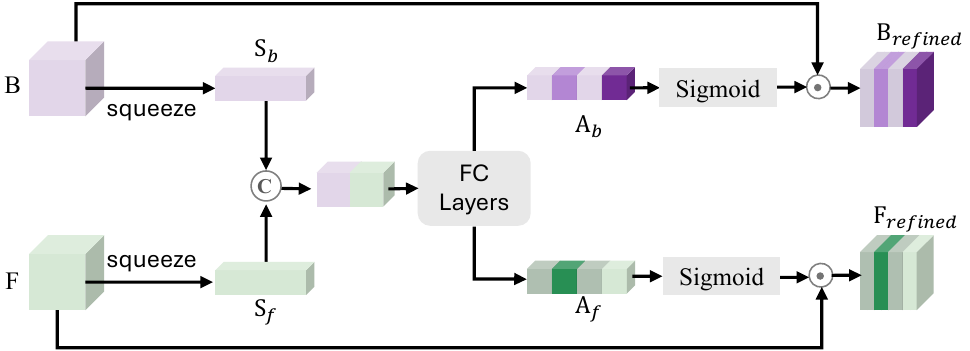}
   \caption{Details of the multimodal transfer module (MMTM) block\replaced{,}{;} adapted from~\cite{joze2020mmtm}.}
   \label{fig:mmtm_module}
\end{figure}

Multimodal Transfer Module (MMTM) architecture~\cite{joze2020mmtm} is equipped with a fusion branch, in which a sequence of MMTM blocks inputs features from the two single-modality branches
and, for each block, outputs two weight maps which are multiplied with the original (single-modality) features. The framework is illustrated in Fig.~\ref{fig:mmtm_overview}. The fusion branch does not directly contribute to the final representation. Instead, it is only used to enhance each of the monomodal feature maps in their separate layers. MMTM extends the \textit{squeeze} and \textit{excitation} (SE) module~\cite{hu2018squeeze} to produce weight maps for the channels, as a specific channel attention mechanism. This modified channel attention also receives two sets of single modality features to accommodate multimodal scenarios.

The detailed structure of the MMTM block is shown in Fig.~\ref{fig:mmtm_module}. Feature maps $B$ and $F$ are extracted from each of the monomodal inputs. 
A ``squeeze'' operation compresses the global information within the input feature maps using Global Average Pooling (GAP), resulting in the squeezed features $S_b$ and $S_f$ which are concatenated into a combined multimodal vector. The concatenated vector is passed through two fully connected layers to generate two modulation weight maps ($A_b$ and $A_f$) for the different modalities. The two weight maps are normalized using sigmoid function and used to recalibrate the single-modality branches through a multiplication with the original feature maps in the channel dimension. The MMTM block adapts well across various spatial dimensions and can be easily integrated with existing network structures. However, it has a limited ability to capture interactions between the modalities and primarily adjusts, rather than fully fuses, the features.


\begin{figure*}[t]
  \centering
   \includegraphics[width=1\linewidth]{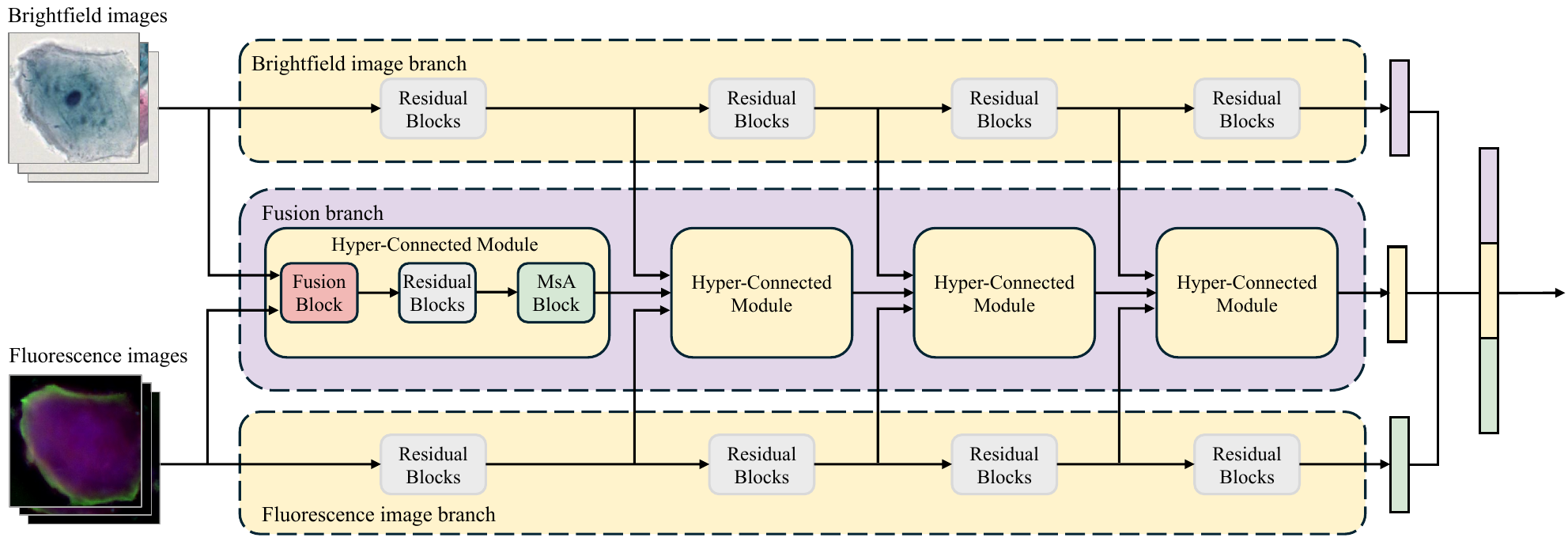}
   \caption{Overview of the hyper-connected convolutional neural network (HcCNN), adapted from on~\cite{bi2020multi}.}
   \label{fig:hccnn_overview}
\end{figure*}

\subsubsection{HcCNN}
\label{subsec:hccnn}

The fusion branch in the Hyper-connected Convolutional Neural Network (HcCNN)~\cite{bi2020multi} receives monomodal features from two single-modality branches but, as opposed to the MMTM network, does not use them to affect the original single-modality branches. Instead, the fused features are directly incorporated as a part of the final representations used for prediction. An overview of HcCNN is depicted in Fig.~\ref{fig:hccnn_overview}. The Hyper-Connected Module (HC module) is used at several levels to perform multimodal feature extraction and fusion. Each HC module comprises a fusion block, a residual block, and a multi-scale attention (MsA) block, to merge and refine information extracted from the different modalities.

\begin{figure}[t]
  \centering
   \includegraphics[width=1\linewidth]{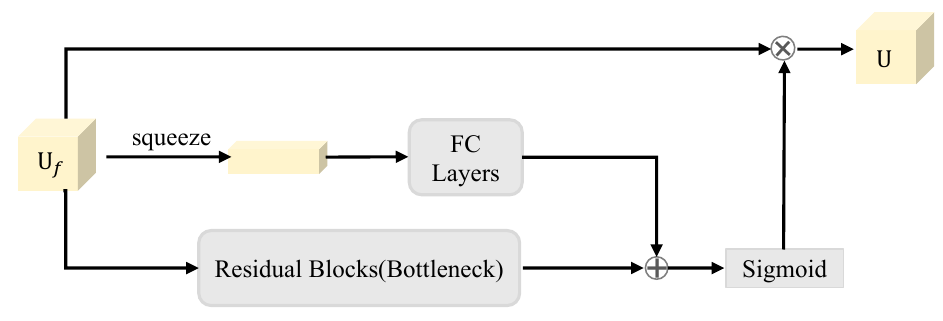}
   \caption{Details of the HcCNN multi-scale attention block (MsA Block), adapted from~\cite{bi2020multi}.}
   \label{fig:hccnn_msablock}
\end{figure}

The fusion block concatenates, as input,  single-modality features $B$ and $F$, and
a fused feature map $U^{'}$ from the previous HC-module. It outputs a fused feature map, which is further processed by several residual blocks, resulting in an enhanced feature set $U_{f}$. Subsequently, the MsA block, illustrated in Fig.~\ref{fig:hccnn_msablock}, is used to refine the complementary information with multi-scale attention. The output $U$ of the MsA block is a spatial weight map, multiplied with the original fused feature $U_{f}$.

The channel-wise attention used in HcCNN to fuse single-modality features adjusts the importance of channels based on global information within each channel, but is not able to fully utilize cross-modal interactions.

\subsubsection{CAFNet}
\label{subsec:cafnet}

\begin{figure*}[t]
  \centering
   \includegraphics[width=1\linewidth]{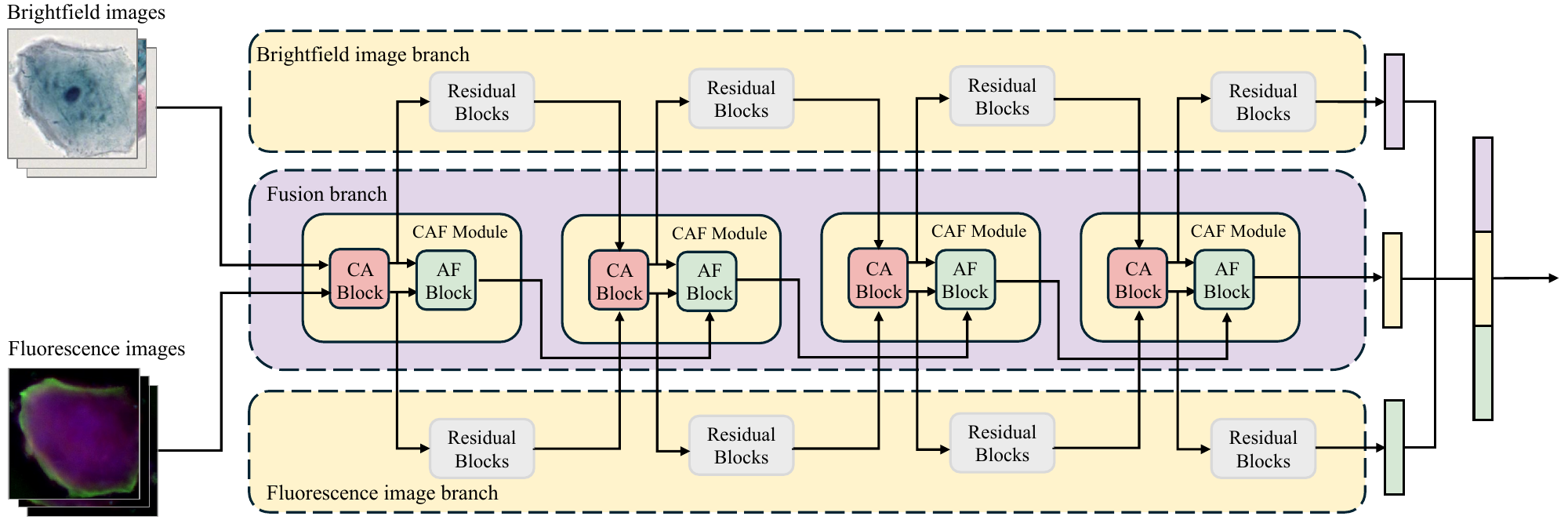}
   \caption{Overview of the Co-Attention Fusion Network (CAFNet) framework, adapted from~\cite{he2023co}.}
   \label{fig:cafnet_overview}
\end{figure*}

\begin{figure}[t]
  \centering
   \includegraphics[width=1\linewidth]{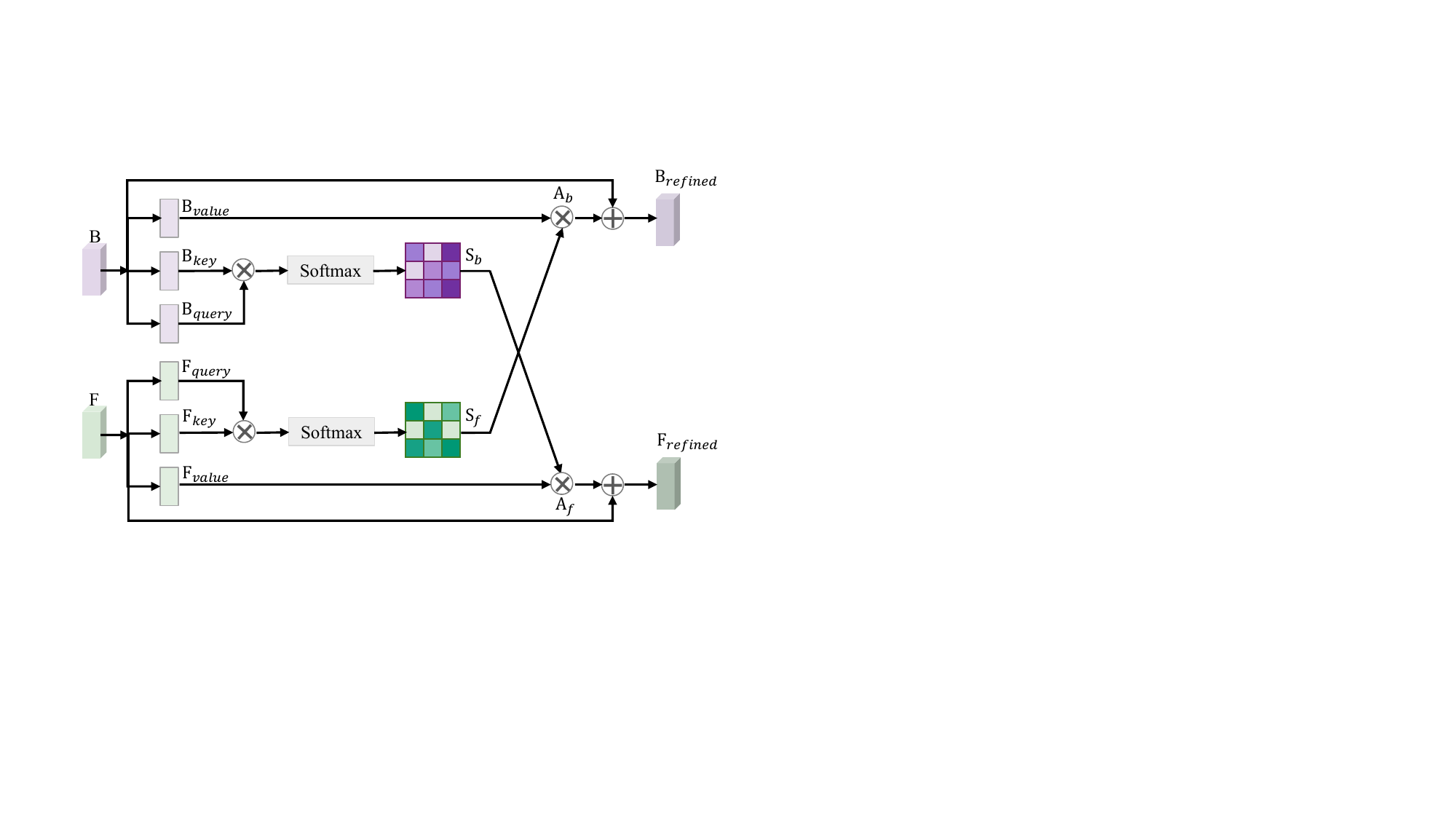}
   \caption{Details of the Co-Attention (CA) block in CAFNet, adapted from~\cite{he2023co}.}
   \label{fig:cafnet_cablock}
\end{figure}

Co-Attention Fusion Network (CAFNet)~\cite{he2023co} draws inspiration from both MMTM and HcCNN. Its fusion branch not only enhances the feature learning of single-modality branches, but also integrates information from different modalities as the fused output. By employing a cross-attention mechanism, CAFNet 
may capture spatial interrelations of the modalities. Fig.~\ref{fig:cafnet_overview} provides an overview of the CAFNet architecture. The fusion branch comprises four CAF modules, each containing a cross-attention (CA) block and an attention fusion (AF) block. The CA block has dual-modality input and output, and employs a cross-attention mechanism to enhance feature representation learning.

Fig.~\ref{fig:cafnet_cablock} illustrates the details of the cross-attention mechanism, which enables the model to 
capture spatial relationships and dependencies between the corresponding regions of the image pairs.
Three metrics are learned for each modality: ``key'' and ``query'' are used to generate weight maps, which are multiplied with the ``value'' metrics of the other modality to support multimodal interaction and improve the feature learning.  

\begin{figure*}[t]
  \centering
   \includegraphics[width=1\linewidth]{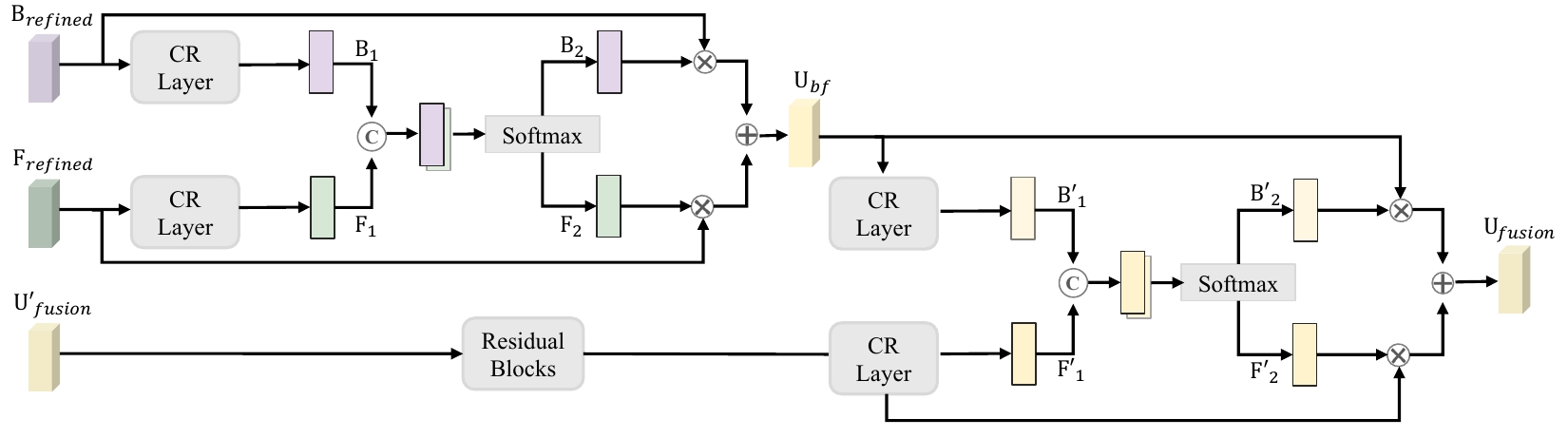}
   \caption{Details of the attention fusion block (AF Block), adapted from~\cite{he2023co}. ``CR Layer'' stands for Channel Reduction, which uses three consecutive convolutions to produce a 1-channel feature map (e.g., $B_1$, $F_1$, $B^{'}_1$, $F^{'}_1$). Corresponding 1-channel features are concatenated to form fusion feature maps, which are then processed through a softmax function to generate weight maps for pixel-wise attention fusion.}
   \label{fig:cafnet_afblock}
\end{figure*}

The AF block, illustrated in Fig.~\ref{fig:cafnet_afblock}, introduces a pixel-wise attention mechanism to dynamically tune weight ratios for multimodal feature fusion.
It takes as input refined features $B_{\text{refined}}$ and $F_{\text{refined}}$ from the CA block, fuses them (pixel-wise attention) into  $U_{bf}$, and then (by the same mechanism), fuses $U_{bf}$ with $U^{'}_{\text{fusion}}$ from the previous stage, to obtain the final feature map $U_{\text{fusion}}$.

Finally, the output of the fusion branch is concatenated with the two single-modality feature vectors for the label prediction.

The cross-attention mechanism in CAFNet allows exploiting the interaction between modalities, but increases model complexity, making this model computationally heavy and its training at a higher risk of overfitting. 

\subsection{Task-specific multimodal model adjustments}
\label{subsec:modifications}

Our aim is to evaluate if multimodal information fusion of BF and FL image data can contribute to improve OC detection from cytological WSI, compared to what is achieved based on either of the two modalities (BF and FL) alone. We therefore compare performances of monomodal, as well as several multimodal approaches (described above) on our created BF \& FL multimodal OC dataset. 

For a meaningful and fair method comparison, \added{we implement} suitable modifications 
\deleted{are implemented} across all \added{the} considered fusion methods. In particular, the original versions of the HcCNN~\cite{bi2020multi} and \replaced{CAFNet~\cite{he2023co}}{MMTM~\cite{joze2020mmtm}} methods employ \deleted{a}
loss function\added{s} designed for the multi-label classification tasks targeted in the original publications\replaced{.}{,} \replaced{HcCNN uses}{and use} a single loss function to evaluate features concatenated from three branches against ground truth labels. In contrast, the original CAFNet\deleted{~\cite{he2023co}} is based on a more intricate (and computationally costly) loss function architecture, with each of the three branches subjected to an individual loss function, supplemented by an additional overall loss function for the concatenated output from all branches, resulting in four collaborative loss functions within its framework. \added{As for MMTM~\cite{joze2020mmtm}, it is originally proposed for tasks unrelated to biomedical applications, with no details provided for the loss functions in the original publication.} \added{To better accommodate our single-label binary classification task and the use of mixup data augmentation, } \replaced{we}{We} adjust the considered intermediate fusion methods by utilizing the same \added{mixup-based} binary classification loss on the final prediction for all the approaches evaluated in this study. 



To fit our OC dataset, we slightly modify the input convolution of all approaches with the correct channel count for each modality, i.e., 3 channels for BF image input and 4 channels for FL image input. For early fusion, we concatenate the channels into a multimodal image with 7 channels, which are processed by a single-branch network.

We use ImageNet~\cite{russakovsky2015imagenet} pretrained ResNet-50 as the backbone for all networks, except the first and last layers which are trained from scratch (randomly initialized). 
The multi-branch networks use the same architecture for each branch, but learn different parameters.

\section{Experimental Setup}
\label{sec:experiments}

\subsection{Data partitioning}
\label{subsec:data_setup}

We utilize a variation of 3-fold cross-validation (CV) to efficiently exploit our limited data.
To prevent data leakage, we partition the dataset into four subsets with individual patients. Each partition contains two patients diagnosed with oral cancer and two or three healthy (i.e., non-oral cancer) individuals. Table~\ref{tab:group} provides a summary of the dataset partitioning, detailing the numbers of images and patients from cancer/healthy groups across the four partitions. Within each partition, the cancer and non-cancer ratios are nearly identical, with the proportion of cancer images consistently around 22\% in all partitions, reflecting the overall distribution.

We designate partition 0 as a 
validation set for choosing suitable hyperparameters, while the remaining three partitions are utilized in a rotating manner for testing and training in each fold. 
To not waste useful training data, our evaluation is divided into two distinct phases. In the ``Initial Validation'' phase, we first tune hyperparameters using partition 0 as a common validation set for each fold. Following this, we move into the ``Full Training'' phase, rejoining  partition 0 into the training set and retrain the network (using the tuned set of hyperparameters) for the final evaluation on the (for the given fold) strictly separated test set.

        

\begin{figure}[t]
  \captionof{table}{Summary of the OC dataset: number of patches and patients, for each class and each partition.  }
  \label{tab:group}
  \centering
  \resizebox{1.\linewidth}{!}{
  \begin{tabular}{@{\,}cc@{\;\;\;}c@{\;\;\;}c@{\,}}
    \toprule
     & Cancer & Healthy & Total \\ 
    Partition & patches (patients) & patches (patients) & patches (patients) \\
    \midrule
    0 & 40,764 (2) & 149,796 (3) & 190,560 (5)\\
    1 & 39,605 (2) & 144,914 (3) & 184,519 (5)\\
    2 & 42,523 (2) & 147,769 (2) & 190,292 (4)\\
    3 & 44,506 (2) & 156,688 (3) & 201,194 (5)\\
    \midrule
    Sum & 167,398 (8) & 599,167 (11) & 766,565 (19) \\
    \bottomrule
  \end{tabular}}
\end{figure}


\subsection{Data augmentation} 
BF and FL provide distinct imaging modalities with major differences in image appearance and characteristics.
For efficient learning, it is essential to implement modality-specific data augmentation strategies.
For the BF images, we apply the following data augmentation steps: either (probability $p=0.4$)
posterization to 3-bit color depth, or ($p=0.2$)
Gaussian blurring (kernel size 5 and sigma 1.5),  or 
($p=0.4$) solarization, inverting pixel values above a threshold of 100. Subsequent color jittering changes brightness by up to 0.5, and contrast, saturation, and hue by up to 0.2 each. Following the jittering, we normalize BF images based on the mean and variance of all BF images in the dataset (this step does not require access to any labels). 
For FL images, we individually apply color jittering to each channel (brightness and contrast at 0.8), along with Gaussian blurring (kernel size 5, sigma in the range 0.3 to 3.2). These images are then normalized using the mean and variance from all FL images. Finally, geometric transforms, such as random flipping and \replaced{resizing}{cropping} are jointly applied to both BF and FL images, ensuring the preservation of the spatial alignment of the two modalities during the data augmentation step. \added{Specifically,} we resize (bilinear interpolation) the 256 $\times$ 256 images to a uniform size of 224 $\times$ 224 to fit a joint standard size of the architectures. Mixup data augmentation is then used to create mixed (joint BF and FL) inputs by linearly combining each sample with another randomly chosen sample from the same batch~\cite{zhang2018mixup}, using $\alpha = 0.8$. 
For validation and testing, we exclusively apply the same normalization and resizing as in training (i.e., no test-time augmentation).

\subsection{Training details}
\label{subsec:details}

The total number of training epochs is set to 30, selected based on performance stability on the validation set (observed for all the models), with the last epoch used for final evaluation.  The batch size is set to 256 for all methods, except CAFNet for which we use a batch size of 128 due to its higher memory requirements. We adopt the AdamW optimizer with an initial learning rate $8 \times 10^{-5}$. The weight decay is set to 0.1. The \textit{CosineAnnealing} learning rate scheduler is used, with minimum learning rate set to $1 \times 10^{-7}$ and  T\_max set to the number of training epochs, which is 30 in our case. We use a \added{mixup-based} binary cross-entropy \added{(BCE)} loss with class weights\deleted{ 22\% for non-cancer cases and 78\% for cancer cases}. \added{Specifically, for each training batch, samples enhanced with the mixup technique are given by}
\begin{equation}
\added{  x_\text{mix} = \lambda \cdot x_a + (1 - \lambda) \cdot x_b,  }
\end{equation}
\added{where $\lambda \sim \text{Beta}(\alpha, \alpha)$ is a coefficient that controls the mixing proportion of the original data $x_a$ and the perturbed data $x_b$. The model’s output logits are defined as $\hat{y}=p_\theta(x_\text{mix})$, where $\theta$ is the network parameter, then the mixup-based BCE loss is}
\begin{equation}
\added{  \mathcal{L} = \lambda \cdot \text{BCE}(\hat{y}, y_a, w_a) + (1 - \lambda) \cdot \text{BCE}(\hat{y}, y_b, w_b),  }
\end{equation} 
\added{where $y_a$ and $y_b$ denote the labels corresponding to $x_a$ and $x_b$, respectively. The weights $w_a$ and $w_b$ are used to adjust imbalanced classes, with a value of 0.22 for non-cancer cases and 0.78 for cancer cases, both are computed from the training dataset.}

All hyperparameters mentioned above are selected based on observed performance during the ``Initial Validation'' phase. Each parameter is evaluated across a range of values. The final choice is determined by the highest average F1 score achieved. 
 All methods are implemented using the PyTorch framework and trained on a single NVIDIA A100 Tensor Core GPU with 80GB memory.

\subsection{Comparison and evaluation metrics} 
\label{subsec:comparison_metric}

The compared methods are divided into three categories: (i) single-modality networks; (ii) simple multimodal methods (i.e., early and late fusion); and (iii) intermediate fusion methods (MMTM~\cite{joze2020mmtm}, HcCNN~\cite{bi2020multi}, and CAFNet~\cite{he2023co}). 
We compare their performance utilizing the following metrics: F1 score, accuracy, Receiver Operating Characteristic Area Under the Curve (ROC AUC), recall, and precision. We also present the full confusion matrix for a comprehensive results' view. Due to the imbalanced nature of our OC dataset, we select F1 score as the primary evaluation metric, while also reporting other metrics for completeness.

\begin{figure}[t]
  \captionof{table}{Average F1 \replaced{score}{Score}, Accuracy, ROC AUC,  Recall, and Precision under 3-fold cross\added{-}validation on the OC \replaced{dataset}{Dataset}, comparing data augmentation with and without color jitter for the monomodal BF and FL image\deleted{s}-only methods.}
  \label{tab:nojitter}
  \centering
  \resizebox{1.\linewidth}{!}{
  \begin{tabular}{@{}lc@{\;\;\;}c@{\;\;\;}c@{\;\;\;}c@{\;\;\;}c@{}}
    \toprule
    Method & F1 score & Accuracy & ROC AUC & Recall & Precision \\
    \midrule
    BF-only & 0.6944 & 0.8521 & 0.8958 & 0.7265 & 0.6894 \\
    BF-only w\kern-0.1em /\kern-0.1em o color jitter &0.6776 & 0.8482 & 0.8779 & 0.7529 & 0.6563  \\
    \midrule
    FL-only& 0.7399 & 0.8768 & 0.9039 & 0.7736 & 0.7246 \\
    FL-only w\kern-0.1em /\kern-0.1em o color jitter &0.5514 & 0.7336 & 0.7404 & 0.6481 & 0.5196  \\
    \bottomrule
  \end{tabular}}
\end{figure}

\section{Results and Analysis}
\label{sec:results}

\subsection{Brightfield vs. Fluorescence}
\label{subsec:BFvFL}
We first compare the performance of single-modality methods (i.e., BF-only and FL-only). As summarized in Table~\ref{tab:nojitter} (and also Table~\ref{tab:all}), the FL-only method outperforms the BF-only method \replaced{\wrt}{for} all considered metrics, in particular showing 4.55 percentage points of improvement in terms of the average F1 score (under 3-fold cross-validation). 
Consistently, Table~\ref{tab:confu} reports higher counts of True Negatives (TN) and True Positives (TP) for the FL method. 

We perform an ablation study regarding the used pre-processing steps and observe that the color jittering is the most important, particularly for the FL image augmentation. 
Table~\ref{tab:nojitter} presents the performance of BF-only and FL-only methods, with and without color jitter in the data augmentation processes. 
The methods utilizing the complete data augmentation strategy consistently outperform those where color jitter is omitted. We therefore conduct all further experiments utilizing the complete augmentation.

\subsection{Monomodal vs. Multimodal}
\label{subsec:1v2}
\begin{figure*}[t]
    \centering
    \begin{subfigure}[b]{0.32\textwidth}
        \includegraphics[width=\linewidth]{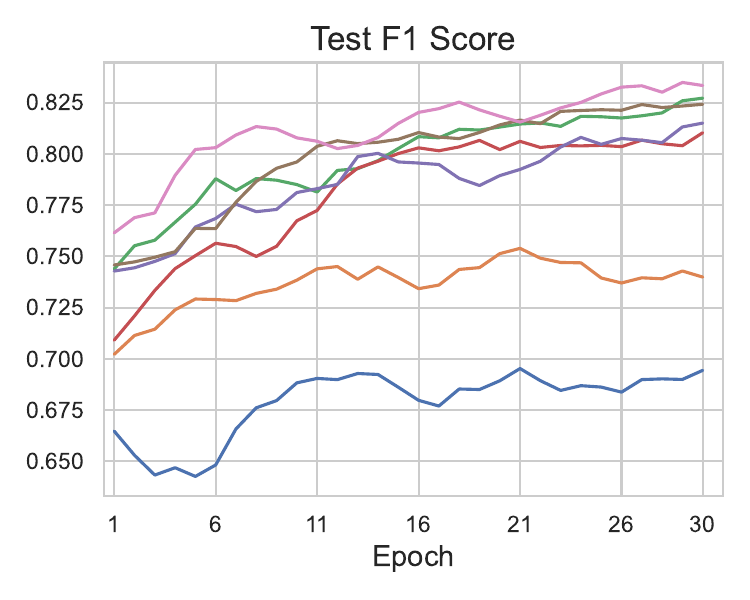}
        \caption{F1 score curve across epochs.}
        \label{fig:curve_f1}
    \end{subfigure}
    \hfill
    \begin{subfigure}[b]{0.32\textwidth}
        \includegraphics[width=\linewidth]{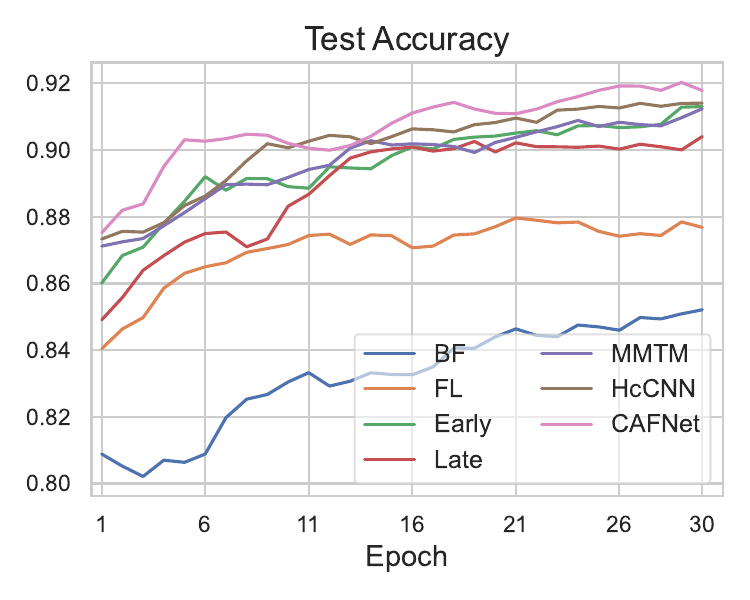}
        \caption{Accuracy curve across epochs.}
        \label{fig:curve_acc}
    \end{subfigure}
    \hfill 
    \begin{subfigure}[b]{0.32\textwidth}
        \includegraphics[width=\linewidth]{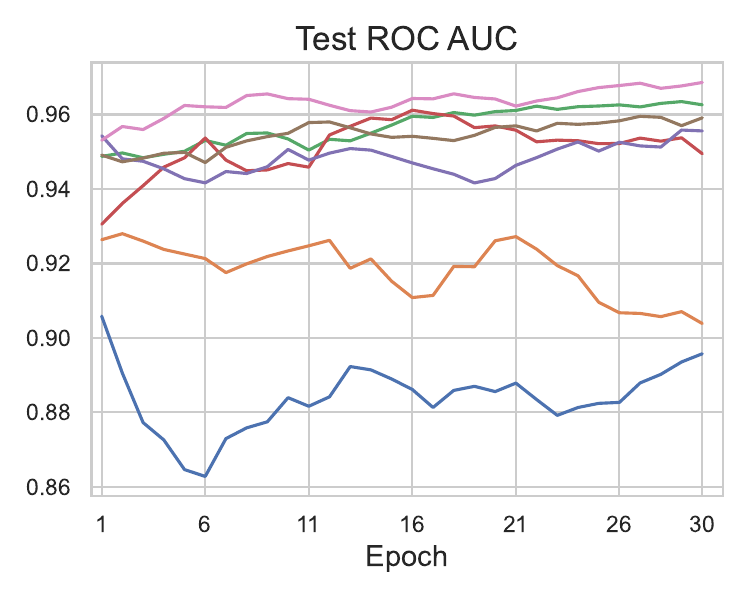}
        \caption{ROC AUC score curve across epochs.}
        \label{fig:curve_rocauc}
    \end{subfigure}
    \caption{\replaced{Test set p}{P}erformance metrics for the considered monomodal and multimodal classification \replaced{methods}{approaches} on the OC \replaced{d}{D}ataset, averaged over 3-fold cross-validation, for each epoch during (the final) training. The legend is consisten\replaced{t}{d} for all three plots.} 
    \label{fig:metrics}
\end{figure*}

We conduct an overall comparison between multimodal approaches and single-modality techniques. The results, presented in Table~\ref{tab:all}, show that multimodal methods consistently outperform single-modality methods over all metrics. (This trend is confirmed over all folds, as presented in Table~\ref{tab:f1_scores}-\ref{tab:precision} in the Appendix.) For example, early fusion reaches an average F1 score improvement of over 8 percentage points compared to the best-performing single-modality method, FL-only. The same trend extends to accuracy, ROC AUC, precision and recall, where multimodal approaches also excel. Table~\ref{tab:confu} provides further evidence, showing higher counts of TN and TP, and greater accuracy rates for multimodal methods. 
Fig.~\ref{fig:metrics} visually confirms the consistent superiority of multimodal approaches throughout the training process. This continuous and consistent advantage underscores the inherent benefits of utilizing multiple modalities.

\begin{figure*}[t]
  \captionof{table}{Performance of the considered methods under 3-fold cross-validation: Average \textbf{F1 score}, \textbf{Accuracy},  \textbf{ROC AUC},  \textbf{Recall}, and \textbf{Precision} (with standard deviations). Higher scores indicate better performance. The highest averages are highlighted in bold. }
  \label{tab:all}
  \centering
  \resizebox{0.9\linewidth}{!}{
  \begin{tabular}{@{}lccccc@{}}
    \toprule
    Method & F1 score $\pm$ std &  Accuracy $\pm$ std & ROC AUC $\pm$ std & Recall $\pm$ std & Precision $\pm$ std\\
    \midrule
    BF-only & $0.6944 \pm 0.1267$& $0.8521 \pm 0.0735$&$0.8958 \pm 0.0660$&$0.7265 \pm 0.0539$&$0.6894 \pm 0.2443$ \\
    
    FL-only &$0.7399 \pm 0.0817$& $0.8768 \pm 0.0494$&$0.9039 \pm 0.0507$&$0.7736 \pm 0.0208$&$0.7246 \pm 0.1750$ \\
    \midrule
    Early &$0.8273 \pm 0.0959$& $0.9130\pm0.0525$&$0.9626\pm0.0397$&$\textbf{0.9192} \pm 0.0665$&$0.7619\pm 0.1551$ \\
    
    Late&$0.8104 \pm 0.1005$& $0.9041 \pm 0.0585$&$0.9495 \pm 0.0337$&$0.8893 \pm 0.0267$&$0.7593 \pm 0.1845$\\
    \midrule
    MMTM&$0.8151\pm 0.0760$&  $0.9124 \pm 0.0425$&$0.9556 \pm 0.0245$&$0.8541 \pm 0.0692$&$\textbf{0.7978} \pm 0.1749$\\
    
    HcCNN &$0.8243 \pm 0.1006$ & $0.9141 \pm 0.0546$&$0.9591 \pm 0.0346$&$0.8797 \pm 0.0410$&$0.7887 \pm 0.1820$\\
    
    CAFNet&$\textbf{0.8334} \pm 0.0954$ & $\textbf{0.9179} \pm 0.0529$&$\textbf{0.9686} \pm 0.0351$&$0.8994 \pm 0.0756$&$0.7934 \pm 0.1819$\\
    \bottomrule
  \end{tabular}}
\end{figure*}

\begin{figure}[t]
  \captionof{table}{The overall results of the \textbf{confusion matrix} for each method, showing the numbers of True Negatives (TN), False Positives (FP), False Negatives (FN), and True Positives (TP). The highest values for TN and TP are highlighted in bold, corresponding to the lowest values for FP and FN, respectively.}
  \label{tab:confu}
  \centering
  \resizebox{.9\linewidth}{!}{
  \begin{tabular}{@{}lcccc@{}}
    \toprule
    Method & TN$\uparrow$ & FP$\downarrow$  & FN$\downarrow$  & TP$\uparrow$ \\
    \midrule
    BF-only & 398,684	&50,687	&34,887	&91,747 \\
    FL-only & 407,268	&42,103	&28,759	&97,875 \\
    \midrule
    Early & 409,638	&39,733	&\textbf{9,932}	&\textbf{116,702} \\
    Late &408,422& 40,949& 13,940& 112,694 \\
    \midrule
    MMTM& \textbf{416,293}&\textbf{33,078}&15,086&111,548\\
    HcCNN &415,338&34,033&15,060&111,574\\
    CAFNet& 415,040&34,331&12,483&114,151\\
    \bottomrule
  \end{tabular}}
\end{figure}

    
    

\subsection{Early and Late vs. Intermediate fusion}
\label{subsec:2v2}
As can be seen in Table~\ref{tab:all}, early fusion outperforms late fusion in all observed metrics. Early fusion reaches a 1.69 percentage points higher average F1 score and 0.89 percentage points higher average accuracy than late fusion. Observing Fig.~\ref{fig:metrics} it is clear that early fusion surpasses late fusion in overall performance across our key metrics also throughout the training.

Focusing on the more sophisticated intermediate fusion techniques, 
we observe in Table~\ref{tab:all} and Fig.~\ref{fig:metrics} that MMTM exhibits slightly lower F1 score, accuracy, and ROC AUC compared to the early fusion method. 
HcCNN exceeds MMTM in all metrics and surpasses late and early fusion in accuracy. 
CAFNet emerges as the overall leading method, reaching an F1 score of 83.34\% and the top accuracy and ROC AUC, 
as can be seen in Table~\ref{tab:all}. 
Table~\ref{tab:confu} reveals that MMTM records the highest number of TN and early fusion captures the most TP, while CAFNet effectively balances the number of TN and TP. 


\subsection{Additional experiments}
\label{subsec:diss}

\begin{figure*}[t]
    \centering   
    \begin{subfigure}[b]{0.32\textwidth}
        \includegraphics[width=\linewidth]{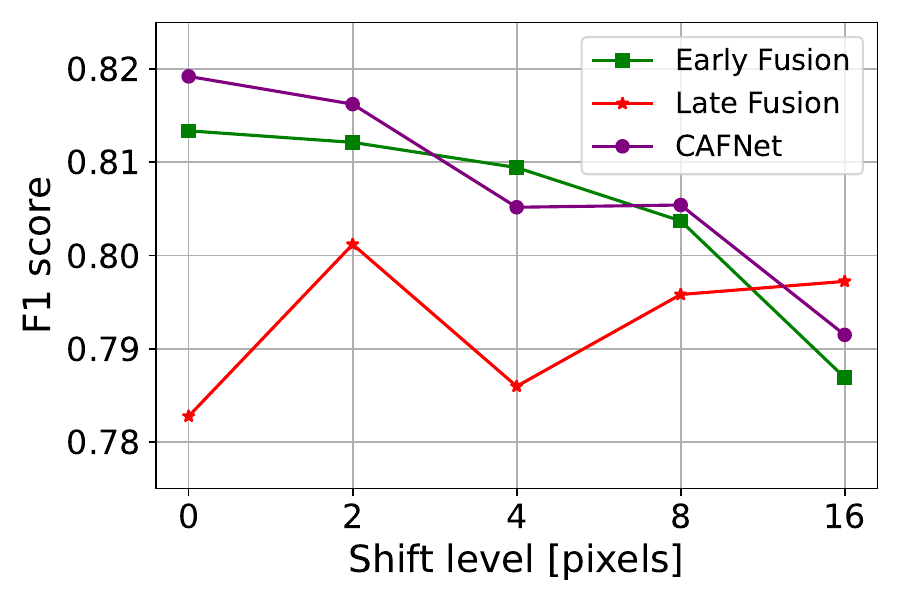}
        \caption{F1 score across different shift levels.}
        \label{fig:f1_shift}
    \end{subfigure}
\qquad
    \begin{subfigure}[b]{0.32\textwidth}
        \includegraphics[width=\linewidth]{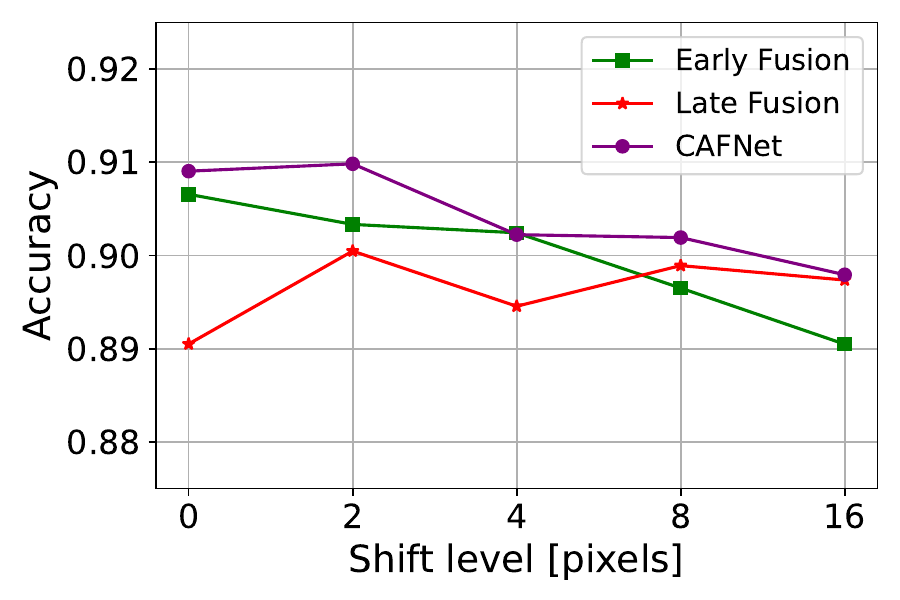}
        \caption{Accuracy across different shift levels.}
        \label{fig:acc_shift}
    \end{subfigure}
    \caption{Performance metrics for early fusion, late fusion and CAFNet methods at increasing levels of displacement between the two modalities (emulating a less accurate image registration).} 
    \label{fig:noalign}
\end{figure*}

\vspace{0.1in}
\noindent\textbf{Impact of misalignment.\,}
We explore the impact of misalignment of (the multimodal) input images on the multimodal fusion performance, with a specific focus on early, late, and CAFNet methods. To introduce alignment discrepancies to multimodal inputs, we manually shift the BF modality images by $d \in \{0,\added{2},4,8,16\}$ pixels, while the FL images are kept at their original position. 
The results are presented in~Fig.~\ref{fig:noalign}. We observe that the performances of \replaced{early fusion and CAFNet}{all three methods} degrade \deleted{sharply} across \replaced{both}{all} metrics as the misalignment 
is introduced. 
As expected, late fusion appears the most robust \wrt variations in alignment, and is the method which reaches the highest F1 score \deleted{and accuracy} 
for the \added{more severely} misaligned data. 
CAFNet, which works best when images are well aligned, experiences a \replaced{substantial}{dramatic} drop in performance if the multimodal images are misaligned. The pixel-wise cross-modal attention of the method appears to give a clear performance boost on well aligned data, however this boost breaks down once faced with misaligned data. 
A general observation is that accurate alignment of the multimodal data appears to be crucial for reaching top performance.

\vspace{0.1in}
\noindent\textbf{Attentions in multimodal fusion.\,}
He \etal~\cite{he2023co} attribute the state-of-the-art performance of CAFNet to the integration of the CA (Channel Attention) and AF (Attention Fusion) blocks. To quantify the individual contributions of these components, we modify the CAFNet architecture with two variants: CAFNet without CA block and CAFNet without AF block. As presented in Table~\ref{tab:caf_varian}, omitting either the AF block or the CA block decreases the performance over all metrics. Removing the CA block leads to the larger drop in performance, indicating that cross-attention appears to play an important role in the multimodal information fusion.


\begin{figure}[t]
  \captionof{table}{Performance of CAFNet and its variants (CAFNet without the CA block and CAFNet without the AF block) on the OC dataset in terms of F1 score, accuracy, ROC AUC score, recall and precision.}
  \label{tab:caf_varian}
  \centering
  \resizebox{1.\linewidth}{!}{
  \begin{tabular}{@{}lc@{\;\;\;}c@{\;\;\;}c@{\;\;\;}c@{\;\;\;}c@{}}
    \toprule
    Method & F1 score & Accuracy & ROC AUC & Recall & Precision \\
    \midrule
    CAFNet & 0.8334 & 0.9179 & 0.9686 & 0.8994 & 0.7934 \\
    CAFNet without AF block  & 0.8252 & 0.9143 & 0.9608 & 0.8844 & 0.7881 \\
    CAFNet without CA block & 0.7910 & 0.8921 & 0.9448 & 0.8813 & 0.7304 \\

    \bottomrule
  \end{tabular}}
\end{figure}

\vspace{0.1in}
\noindent\textbf{Two-stream monomodal methods.\,}
\deleted{In \cite{garcia2019learning}, it is argued that }\replaced{W}{w}hen comparing the performance of a multimodal model and a monomodal model, it is \replaced{relevant}{important} to ascertain that the two models have similar complexity\added{ \cite{garcia2019learning}}. 
\added{We therefore include a performance comparison between a two-stream setup used in the monomodal case (inputting the same modality into both streams) and a two-stream network which processes two modalities.}
\deleted{Therefore, \cite{garcia2019learning} proposes that a two-stream network should be used also in the monomodal when comparing with a two-stream network which processes two modalities, inputting the same modality into both streams.}
Table \ref{tab:2_stream} shows that utilizing two-stream networks for BF and FL modalities slightly outperform their single-stream counterparts. However, these monomodal models still stay far behind the multimodal approaches. For example, the late fusion of similar complexity, with a clear margin provides higher scores across all metrics, highlighting the power of multimodal information fusion.

\begin{figure}[t]
  \captionof{table}{Average F1 score, accuracy, ROC AUC, recall, and precision  under 3-fold cross validation on the OC dataset, comparing single-stream and two-stream networks for BF and FL modalities.}
  \label{tab:2_stream}
  \centering
  \resizebox{1.\linewidth}{!}{
  \begin{tabular}{@{}lccccc@{}}
    \toprule
    Method & F1 score & Accuracy & ROC AUC & Recall & Precision \\
    \midrule
    BF-only & 0.6944 & 0.8521 & 0.8958 & 0.7265 & 0.6894 \\
    two-stream BF-only& 0.7015 & 0.8492 & 0.9005&0.7612&0.6764 \\
    \midrule
    FL-only& 0.7399 & 0.8768 & 0.9039 & 0.7736 & 0.7246 \\
    two-stream FL-only& 0.7451 &0.8792 &0.9048&0.7838& 0.7213 \\
    \midrule
    Late fusion & 0.8104& 0.9041 & 0.9495 & 0.8893 & 0.7593 \\
    \bottomrule
  \end{tabular}}
\end{figure}

\section{Discussion}
\label{sec:discussion}

Our above reported classification results are obtained on the cell/patch level. Considering our overarching goal of improved early OC detection, these results should be aggregated, to provide predictions on the patient level. A simple aggregation strategy is to determine the ratio of the predicted malignant (positive) and healthy (negative) cells for each patient, and to perform dichotomisation of the patients to the two classes based on, e.g., a suitably determined threshold. 

\begin{figure}[t]
    \centering
    \begin{subfigure}[b]{0.49\textwidth}
        \includegraphics[width=\linewidth,trim= 0 7mm 0 0,clip]{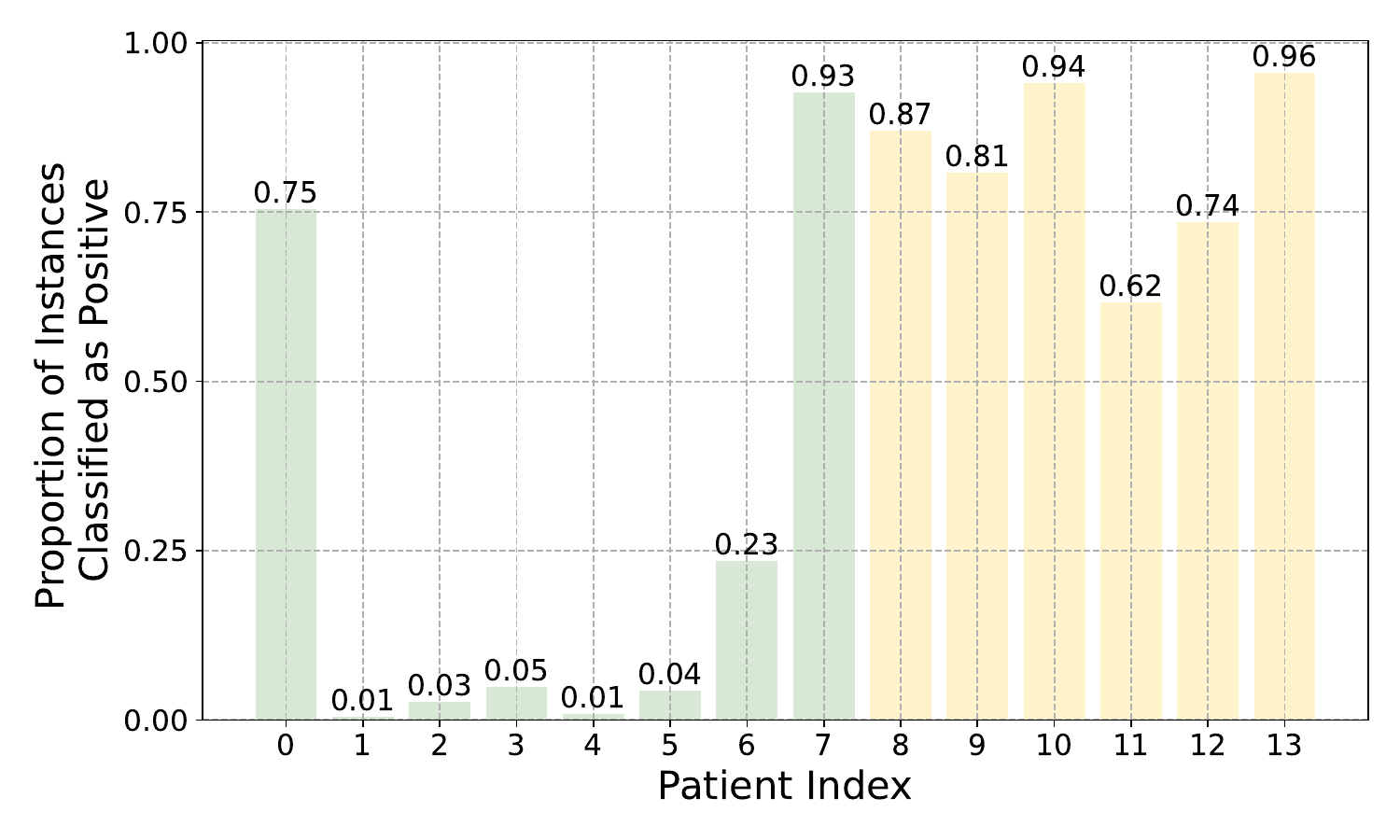}
        \caption{BF-only.}
        \label{fig:bf_patient}
    \end{subfigure}
    \hfill 
    \begin{subfigure}[b]{0.49\textwidth}
        \includegraphics[width=\linewidth,trim= 0 7mm 0 0,clip]{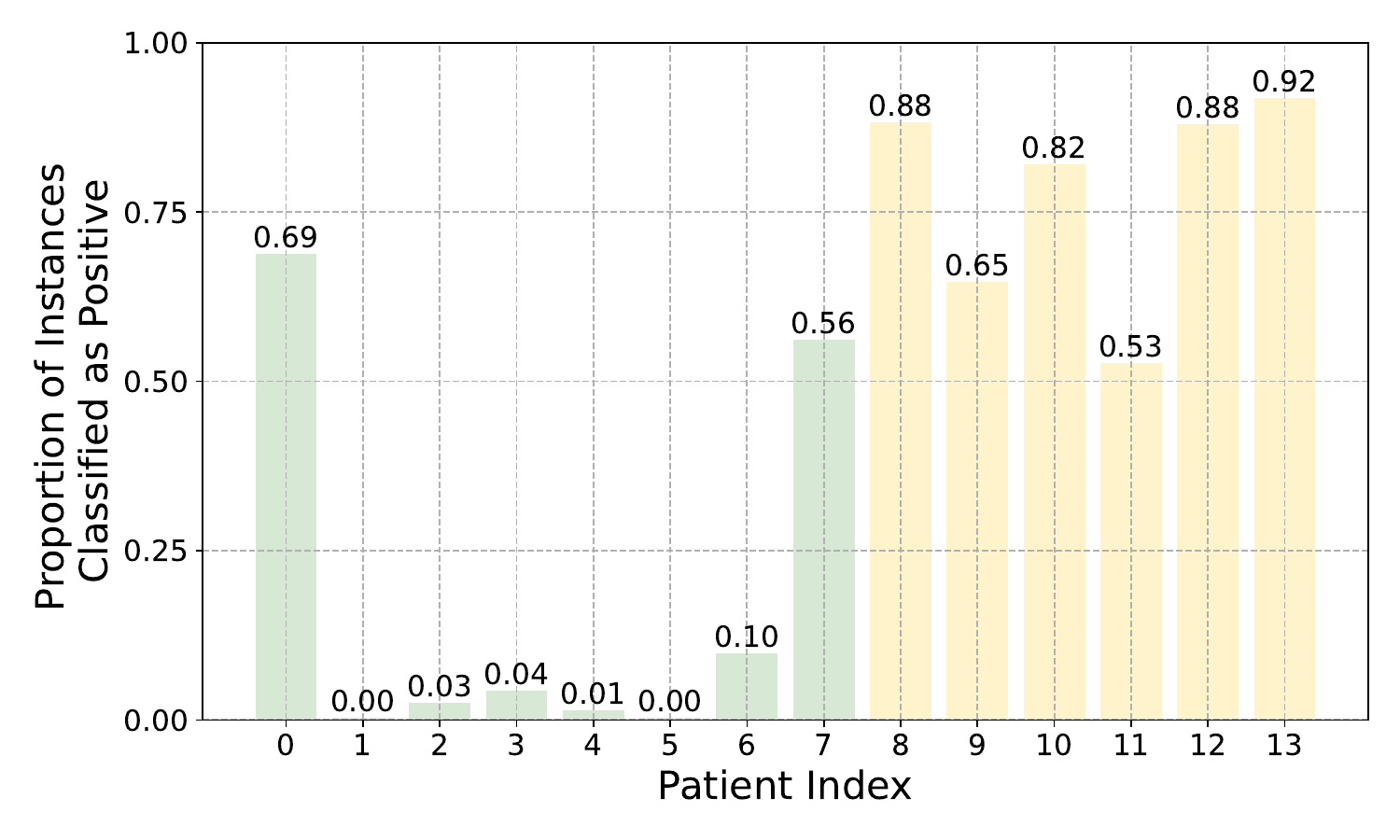}
        \caption{FL-only.}
        \label{fig:fl_patient}
    \end{subfigure}
    \hfill
    \begin{subfigure}[b]{0.49\textwidth}
        \includegraphics[width=\linewidth,trim= 0 7mm 0 0,clip]{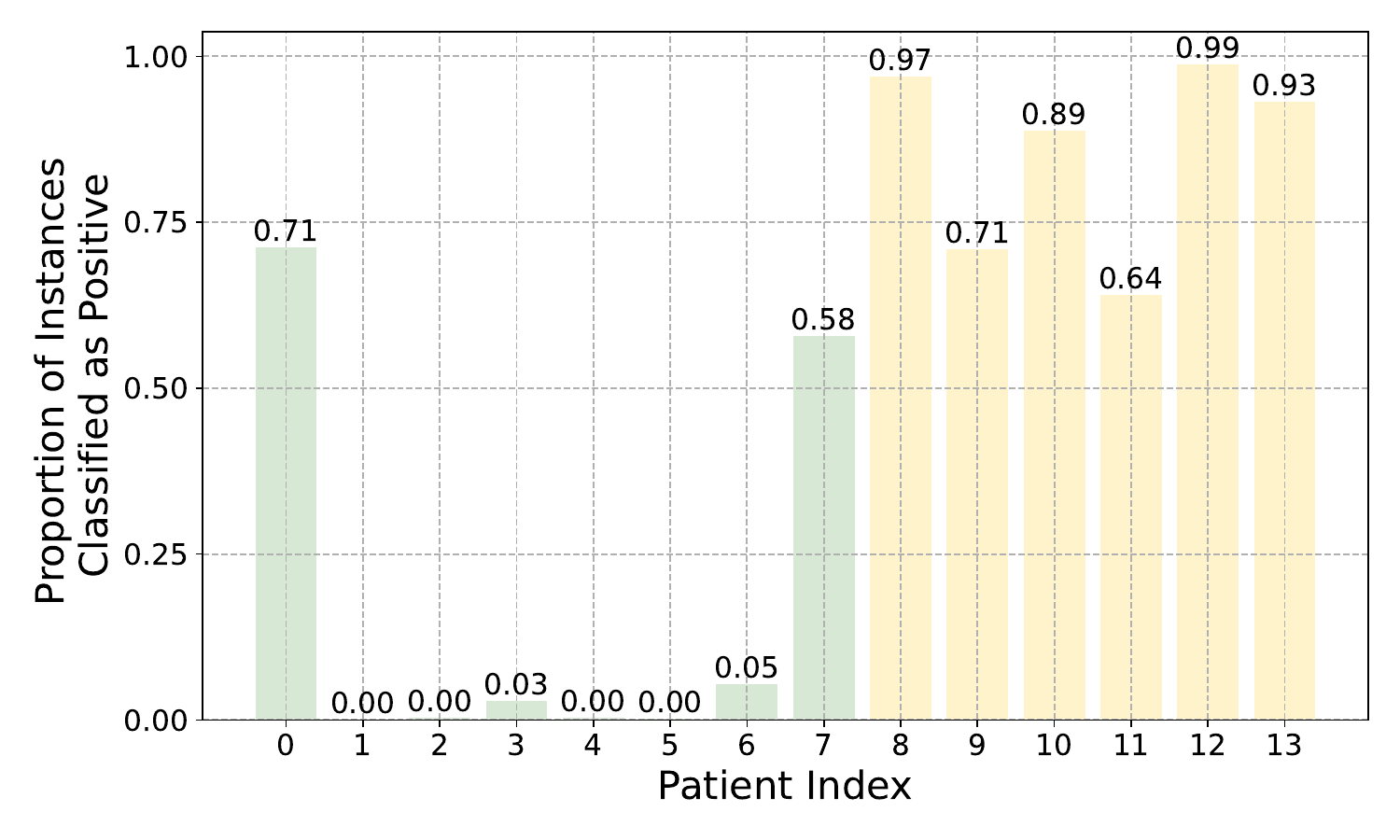}
        \caption{CAFNet.}
        \label{fig:caf_patient}
    \end{subfigure}
    \caption{The relative numbers of cells predicted as positive for each patient in the oral cancer dataset, using different methods. Instances with a sigmoid score $>0.5$ are classified as positive, indicating the presence of oral cancer. Patients 0--7 (green) are negative, and the rest (yellow) are positive.} 
    \label{fig:patient_level}
\end{figure}

 In Fig.~\ref{fig:patient_level} we present the relative numbers of cells predicted as positive, by the BF-only, FL-only, and CAFNet models. A patient-level decision should be reached taking into consideration more information, but a simple observation of ratios does allow some observations.
By placing the threshold for a positive/negative patient-level classification at 
60\% of positive cells, we obtain, by using CAFNet model, classification of patients with 0 FN and 1 FP (with the corresponding F1 score: 0.92; accuracy: 0.93; recall: 1.0; precision: 0.86). 
However, even if these results are highly promissing, the so far acquired dataset (19 patients, out of which 5 are excluded in this evaluation to prevent data leakage) is too small for any reliable conclusions to be yet drawn at the patient level. We are continuously acquiring more (multimodal) data, and based on the results of this study, we expect that FM imaging will be a part of our future pipeline.

Considering the general challenge of multimodal image alignment, of our particular interest was evaluating the effect of misaligned multimodal input on the performance of the different information fusion models/strategies. Our observations \deleted{to a large extent} support the often reported findings that the alignment of the input data has a strong impact on performance of the early fusion strategies. However, we also observe a similar effect in case of \replaced{the}{late and} intermediate fusion approach\deleted{es}. \deleted{This behavior, in particular of the late fusion deserves further exploration;} \added{It is reasonable to assume that,} in case of CAFNet -- the best performing approach on the aligned data\replaced{,}{ -- it is reasonable to assume that} the sophisticated pixel-wise attention mechanisms responsible for its excellent performance on the aligned data, are also a main reason for a high sensitivity of the model to misalignment in the multimodal input.  

The attention mechanisms and the extra fusion branch added to the intermediate model architectures  substantially increase memory and computational demands, which may limit their use in resource-constrained environments. Our experiments show that early fusion, requiring only about one-third of the training time of CAFNet 
achieves comparable results and can be a recommended approach to balance the advantages of multimodal information with the required resources for its processing.    

\section{Conclusion}
\label{sec:conclusions}
In this study, we present a workflow for the early detection of oral cancer from cytology WSIs, exploiting the information gain reached by a multimodal approach.
We create a multimodal OC \replaced{d}{D}ataset, imaged with BF and FL microscopy, and perfectly aligned on the patch/cell level.  
We evaluate various deep learning strategies for fusing multimodal information from the created dataset. Our findings demonstrate that multimodal methods consistently outperform single-modality approaches across all evaluated metrics. Among the methods investigated, CAFNet -- an intermediate fusion approach -- stands out as the most effective, achieving an F1 score of 83.34\% and an accuracy of 91.79\% for the cell-level classification.

\section*{Acknowledgements}
This work is supported by: Sweden’s Innovation Agency (VINNOVA), grants 2017-02447, 2020-03611, and 2021-01420, the Swedish Research Council, grants 2017-04385 and 2022-03580, and Cancerfonden projects 22 2353 Pj and 22 2357 Pj. A part of the computations was enabled by resources provided by the National Academic Infrastructure for Supercomputing in Sweden (NAISS) and the Swedish National Infrastructure for Computing (SNIC) at Chalmers Centre for Computational Science and Engineering (C3SE), partially funded by the Swedish Research Council through grants 2022-06725 and 2018-05973.
We are grateful to M.~Mati\'c for accurate cell location annotations.

\section*{Additional information}
The study was performed in compliance with the Declaration of Helsinki and approved by the Ethical Review Board Stockholm Sweden (2015-1213-31 and 2019-00349). Informed written consent was obtained from all participants. The study does not involve minors.

\subsection*{Competing interests}
The authors declare no competing interests.




\printcredits

\bibliographystyle{cas-model2-names}

\bibliography{main}

\bio{}
\endbio


\appendix
\counterwithin{table}{section} 
\counterwithin{figure}{section} 
\section{Appendix}

\begin{figure}[h]
  \captionof{table}{Reached \textbf{F1 scores} for the considered methods under 3-fold cross-validation on the multimodal OC dataset, showcasing each method's performance across three independent folds and their average scores. Higher score indicates better performance, with the highest scores highlighted in bold.}
  \label{tab:f1_scores}
  \centering
  \resizebox{0.7\linewidth}{!}{
  \begin{tabular}{@{}lcccc@{}}
    \toprule
    Method & Fold 1 & Fold 2 & Fold 3 &  Average\\
    \midrule
    BF-only  & 0.6494 & 0.8375 & 0.5963 & 0.6944 \\
    FL-only  & 0.6752 & 0.8318 & 0.7128 & 0.7399 \\
    \midrule
    Early    & 0.7223 & 0.9104 & 0.8492 & 0.8273 \\
    Late     & 0.7091 & 0.9101 & 0.8120 & 0.8104 \\
    \midrule
    MMTM     & \textbf{0.7311} & 0.8792 & 0.8350 & 0.8151 \\
    HcCNN    & 0.7184 & \textbf{0.9185} & 0.8359 & 0.8243 \\
    CAFNet   & 0.7263 & 0.9090 & \textbf{0.8650} & \textbf{0.8334} \\
    \bottomrule
  \end{tabular}}
\end{figure}

\begin{figure}[h]
  \captionof{table}{Reached \textbf{accuracy} for the considered methods under 3-fold cross-validation on the multimodal OC dataset, showcasing each method's performance across three independent folds and their average accuracy. Higher accuracy indicates better performance, with the highest scores highlighted in bold.}
  \label{tab:acc}
  \centering
  \resizebox{0.7\linewidth}{!}{
  \begin{tabular}{@{}lcccc@{}}
    \toprule
    Method & Fold 1 & Fold 2 & Fold 3 &  Average\\
    \midrule
    BF-only & 0.8204 &0.9362 &0.7998 &0.8521 \\
    FL-only & 0.8354 &0.9315 &0.8636 &0.8768 \\
    \midrule
    Early & 0.8567& 0.9606& 0.9219& 0.9130 \\
    Late & 0.8446& 0.9617& 0.9059& 0.9041\\
    \midrule
    MMTM& \textbf{0.8670}& 0.9514& 0.9188& 0.9124\\
    HcCNN & 0.8569& \textbf{0.9658}& 0.9195& 0.9141\\
    CAFNet& 0.8594& 0.9623& \textbf{0.9320}& \textbf{0.9179}\\
    \bottomrule
  \end{tabular}}
\end{figure}

\begin{figure}[h]
  \captionof{table}{Reached \textbf{ROC AUC scores} for the considered methods under 3-fold cross-validation on the multimodal OC dataset, showcasing each method's performance across three independent folds and their average scores. Higher score indicates better performance, with the highest scores highlighted in bold.}
  \label{tab:roc}
  \centering
  \resizebox{0.7\linewidth}{!}{
  \begin{tabular}{@{}lcccc@{}}
    \toprule
    Method & Fold 1 & Fold 2 & Fold 3 &  Average\\
    \midrule
    BF-only & 0.8824&0.9674&0.8374 & 0.8958 \\
    FL-only & 0.8938&0.9589&0.8590& 0.9039 \\
    \midrule
    Early &0.9168&0.9855&0.9856 & 0.9626\\
    Late & 0.9153 &0.9827&0.9504 & 0.9495\\
    \midrule
    MMTM&\textbf{0.9287}&0.9767&0.9614 & 0.9556\\
    HcCNN &0.9210&0.9885&0.9678 &0.9591 \\
    CAFNet& 0.9282 &\textbf{0.9917}&\textbf{0.9860}&\textbf{0.9686}\\
    \bottomrule
  \end{tabular}}
\end{figure}

\begin{figure}[h]
  \captionof{table}{Reached \textbf{recall} for the considered methods under 3-fold cross-validation on the multimodal OC dataset, showcasing each method's performance across three independent folds and their average recall. Higher recall indicates better performance, with the highest scores highlighted in bold.}
  \label{tab:recall}
  \centering
  \resizebox{0.7\linewidth}{!}{
  \begin{tabular}{@{}lcccc@{}}
    \toprule
    Method & Fold 1 & Fold 2 & Fold 3 &  Average\\
    \midrule
    BF-only  & 0.7750 & 0.7360 & 0.6685 & 0.7265 \\
    FL-only  & 0.7972 & 0.7582 & 0.7652 & 0.7736 \\
    \midrule
    Early    & 0.8684 & \textbf{0.8949} & \textbf{0.9944} & \textbf{0.9192} \\
    Late     & \textbf{0.8821} & 0.8669 & 0.9189 & 0.8893 \\
    \midrule
    MMTM     & 0.8424 & 0.7914 & 0.9284 & 0.8541 \\
    HcCNN    & 0.8502 & 0.8621 & 0.9266 & 0.8797 \\
    CAFNet   & 0.8691 & 0.8436 & 0.9854 & 0.8994 \\
    \bottomrule
  \end{tabular}}
\end{figure}

\begin{figure}[h]
  \captionof{table}{Reached \textbf{precision} for the considered methods under 3-fold cross-validation on the multimodal OC dataset, showcasing each method's performance across three independent folds and their average precision. Higher precision indicates better performance, with the highest scores highlighted in bold.}
  \label{tab:precision}
  \centering
  \resizebox{0.7\linewidth}{!}{
  \begin{tabular}{@{}lcccc@{}}
    \toprule
    Method & Fold 1 & Fold 2 & Fold 3 &  Average\\
    \midrule
    BF-only  & 0.5588 & 0.9713 & 0.5382 & 0.6894 \\
    FL-only  & 0.5856 & 0.9211 & 0.6671 & 0.7246 \\
    \midrule
    Early    & 0.6183 & 0.9264 & 0.7409 & 0.7619 \\
    Late     & 0.5928 & 0.9577 & 0.7274 & 0.7593 \\
    \midrule
    MMTM     & \textbf{0.6457} & \textbf{0.9889} & 0.7586 & \textbf{0.7978} \\
    HcCNN    & 0.6220 & 0.9828 & 0.7613 & 0.7887 \\
    CAFNet   & 0.6238 & 0.9855 & \textbf{0.7708} & 0.7934 \\
    \bottomrule
  \end{tabular}}
\end{figure}





\end{document}

%% file: preamble.tex
\PassOptionsToPackage{dvipsnames,svgnames,x11names}{xcolor} 
\usepackage{color,soul} 
\usepackage[markup=underlined, commandnameprefix=ifneeded]{changes}
\usepackage{graphicx}

\usepackage{xspace}
\xspaceaddexceptions{]\}}
\makeatletter
\DeclareRobustCommand\onedot{\futurelet\@let@token\@onedot}
\def\@onedot{\ifx\@let@token.\else.\null\fi\xspace}

\def\wrt{w.r.t\onedot} 
\def\etal{\emph{et al}\onedot}
\makeatother